\documentclass[12pt,draftclsnofoot,onecolumn]{IEEEtran}
\usepackage{cite}
\usepackage{graphicx}
\usepackage{amsfonts}
\usepackage{amssymb}
\usepackage{amsmath}
\usepackage{mathrsfs}
\usepackage{bm}
\usepackage{amsmath}
\usepackage{mdwmath}
\usepackage{mdwtab}
\newtheorem{theorem}{Theorem}
\newtheorem{remark}{Remark}

\begin{document}
\title{Achievable Rates of FDD Massive MIMO Systems with Spatial Channel Correlation}
\author{Zhiyuan Jiang, Andreas F. Molisch,~\IEEEmembership{Fellow,~IEEE}, Giuseppe Caire,~\IEEEmembership{Fellow,~IEEE}, and Zhisheng Niu,~\IEEEmembership{Fellow,~IEEE}
\thanks{Z. Jiang and Z. Niu are with Tsinghua National Laboratory for Information Science and Technology, Tsinghua University, Beijing 100084, China. Emails:
jiang-zy10@mails.tsinghua.edu.cn; niuzhs@tsinghua.edu.cn.

A. F. Molisch and G. Caire are with the Ming Hsieh Department of Electrical Engineering, University of Southern California, Los Angeles, CA 90089-2565, USA. Emails: molisch@usc.edu; caire@usc.edu.

This work is sponsored in part by the National Basic Research Program of China (973 Program: 2012CB316001), the National Science Foundation of China (NSFC) under grant No. 61201191 and No. 61321061, and Hitachi R\&D Headquarter.}}

% make the title area
\maketitle
\vspace{-1cm}
\begin{abstract}
%\boldmath
It is well known that the performance of frequency-division-duplex (FDD) massive MIMO systems with \emph{i.i.d. channels} is disappointing compared with that of time-division-duplex (TDD) systems, due to the prohibitively large overhead for acquiring channel state information at the transmitter (CSIT). In this paper, we investigate the achievable rates of FDD massive MIMO systems with \emph{spatially correlated channels}, considering the CSIT acquisition dimensionality loss, the imperfection of CSIT and the regularized-zero-forcing linear precoder. The achievable rates are optimized by judiciously designing the downlink channel training sequences and user CSIT feedback codebooks, exploiting the multiuser spatial channel correlation. We compare our achievable rates with TDD massive MIMO systems, i.i.d. FDD systems, and the joint spatial division and multiplexing (JSDM) scheme, by deriving the deterministic equivalents of the achievable rates, based on popular channel models. It is shown that, based on the proposed eigenspace channel estimation schemes, the rate-gap between FDD systems and TDD systems is significantly narrowed, even approached under moderate number of base station antennas. Compared to the JSDM scheme, our proposal achieves dimensionality-reduction channel estimation without channel pre-projection, and higher throughput for moderate number of antennas and moderate to large channel coherence time, though at higher computational complexity.
\end{abstract}


\section{Introduction}
Scaling-up multiple-input-multiple-output (MIMO) systems, thus exploiting the spatial degree-of-freedom (DoF), plays a pivotal role in boosting the capacity of next generation wireless communication systems. In cellular systems, it is found desirable to deploy a large number of antennas at base stations (BSs) \cite{Marzetta10}, resulting in what is referred to as the massive MIMO system. Such designs have several advantages, including significant improvements of spectral efficiency and radiated energy efficiency \cite{Ngo13}, immunity to small-scale channel fading due to the channel hardening effect, simplification of the media-access-control (MAC) layer design, etc.

Striving to reap the dramatic throughput gain of massive MIMO systems, it is found that such capacity improvements rely heavily on the availability of channel state information at the transmitter (CSIT). Without CSIT, e.g., when the user channels are identically distributed and are i.i.d. (independent identically distributed) in time/frequency, the total DoF reduces to one \cite{caire03}.\footnote{In such condition it has been shown that even when the CSIT is known within a mean-square error that does not decrease with SNR, the DoF collapses to one \cite{jafar2014}.} In practice, a pilot-assisted CSIT acquisition approach is widely adopted, where the BS first broadcasts downlink channel training sequences, and then listens to the channel feedback from the users. This is the case for the frequency-division-duplex (FDD) system or the uncalibrated time-division-duplex (TDD) system.\footnote{Since in practice TDD reciprocity is quite difficult to obtain, which requires reciprocity calibration of the transmit and receive radio frequency chains. In fact, the only current system that uses MU-MIMO, which is 802.11ac, uses explicit polling of the users through downlink pilots, and explicit quantized closed-loop feedback from the users, even though it is a TDD system.} For the calibrated TDD system, the channel reciprocity is exploited to allow the BS to obtain the CSIT through uplink channel training. Assuming the channel coefficients are i.i.d. for different users and BS antennas, the CSIT acquisition overhead, which leads to a dimensionality loss of the time-frequency resource, scales with the number of BS antennas for FDD systems, and the number of users for TDD systems, respectively. As we scale up the number of BS antennas, the overhead will become prohibitively large for the FDD system. Therefore, it is commonly considered that the TDD mode is the better, if not the only, choice for massive MIMO systems. Nonetheless, since currently deployed cellular systems are dominantly FDD, and many frequency bands are assigned explicitly for use in FDD, it is of great interest to design schemes that realize the massive MIMO gains with an FDD mode.

Given the fact that the dimensionality loss due to CSIT acquisition overhead is devastating with closed-loop channel estimation in FDD and uncalibrated TDD systems, and that the system performance without CSIT is unacceptably poor, it is natural to pose the question whether there exists other information that can be estimated at a much lower cost, while accomplishing the same task as the CSIT. To this end, it is found that the second order channel statistics, specifically the channel correlation matrices (CCMs) of the channel coefficients, are of tremendous help \cite{clerckx08, Adhikary13, Adhikary132, Adhikary14, nam14}. Compared with the instantaneous CSIT realizations, the CCMs, which are determined by user-locations and large-scale fading, vary at a much slower time scale, e.g., seconds to tens of seconds in cellular systems. Therefore, their estimation cost is drastically lower than instantaneous CSIT. In the mean time, recent work shows the CCMs can be leveraged, in many ways, to facilitate FDD massive MIMO transmission. While the optimal transmission scheme with the aid of CCMs is still unclear, significant rate gain can be expected \cite{Adhikary13}.

A large body of work has been done studying TDD massive MIMO systems. The seminal work in \cite{Marzetta10} first proposes to deploy BS antennas with a number much larger than the number of users, eliminating the impact of small-scale channel fading and uncorrelated noise due to the \emph{channel hardening} effect, while only the inter-cell interference remains due to \emph{pilot contamination}. In \cite{Ngo13}, the authors show that in addition to the spectral efficiency improvement, the massive MIMO system increases the radiated energy efficiency by a factor of $M$, where $M$ is the number of BS antennas, or $\sqrt{M}$ in the presence of imperfect channel estimation. Recent work in \cite{muller13} further shows the pilot-contamination problem is \emph{not} inherent. Several other issues are also studied extensively, such as downlink precoding, detection, hardware impairment, etc. \cite{Studer13, bj13, shepard13}.

For FDD massive MIMO systems, the research can be categorized into two directions. One is exploiting the time correlation of the channels, e.g., \cite{Choi14} and references therein, where a trellis-code based quantization codebooks are leveraged to decrease the CSIT estimation overhead. The other is exploiting the spatial correlation of channel coefficients, pioneered by the work in \cite{Adhikary13} and extended in \cite{Adhikary132,Adhikary14,nam14}, which propose the joint spatial division and multiplexing (JSDM) scheme. Based on the JSDM scheme, the users are divided into groups based on their CCMs, and a two-stage precoding is performed, namely the pre-beamforming and the beamforming, which utilize the CCMs to counteract the inter-group-interference and the instantaneous channel estimations to manage the interference inside each group, respectively.

By assuming \emph{perfect} knowledge of the CCMs at both the BS side and the user side, the current work endeavors to optimize the achievable rates of FDD massive MIMO systems. Specifically, we propose eigenspace channel estimation methods to improve the system achievable rates, for the case of spatially correlated channels. The main contributions of this paper include:
\begin{itemize}
\item
The low-rank covariance matrices of the channels are exploited in order to design efficient channel training and feedback schemes, which enable dimensionality reduced channel estimation, e.g., it may suffice to train the downlink broadcast channel (BC) with pilots less than the number of BS antennas. In fact, the proposed channel training and feedback schemes can be seen as an alternative to the pre-projection and effective channel approach in JSDM. We derive deterministic equivalents of the achievable rates for our schemes with a regularized-zero-forcing (RZF) precoder, considering distinct CCMs of different users, the dimensionality loss due to channel training and feedback processes, and the imperfection of channel estimations. The proposed approach requires minimal modifications of the widely-adopted pilot-assisted scheme, thus making it desirable to implement in practice.
\item
The optimal channel training sequences with distinct CCMs for different users is studied for the first time. We propose an iterative algorithm to find the optimal training sequences, within the heuristics of the algorithm, based on maximizing the mutual information between the channel coefficients and the received channel training signals. The training sequences found by the algorithm are shown to improve the system achievable rates substantially, compared with the training sequences optimized for the i.i.d. case.
\item
The Karhunen-Loeve (KL) transform followed by entropy coded scalar quantization (SQ) with reverse water filling bit-loading for the feedback codebook design (KLSQ) is proposed. We compare its performance with two vector quantization (VQ) methods designed for the spatially correlated channel case. It is shown that the KLSQ is a simple way to approach the optimal VQ performance for correlated Gaussian channel vectors. The simplicity is due to the fact that it is only SQ followed by Huffman entropy coding. Therefore, it is of very low complexity for real time implementation, which justifies and motivates its use.
\item
Comprehensive numerical results are given to evaluate the performance. We consider the one-ring channel model and the Laplacian angular spectrum channel model, and compare our achievable sum rate with the TDD system and the i.i.d. FDD system under various system parameters. Significant rate gains are obtained by our proposed channel estimation scheme, in spatially correlated channels. Furthermore, in comparison with the JSDM scheme, it is shown that the achievable sum rate with our proposal is better in most scenarios, except when the channel coherence time is very small and the users are well separated in the angular domain.
\end{itemize}

The remainder of the paper is organized as follows. In Section \ref{sec_SM}, the system model is characterized. In Section \ref{sec_AE}, we specify the proposed eigenspace channel training and feedback schemes, and derive the achievable rates. In Section \ref{sec_DE}, we derive the deterministic equivalents of the achievable rates. Section \ref{sec_nr} gives the simulation results, including the comparison with TDD and i.i.d. FDD systems, and the JSDM scheme, under various system parameters. Finally, in Section \ref{sec_cl}, we conclude our work.

$\mathbf{Notations:}$ Throughout the paper, we use boldface uppercase letters, boldface lowercase letters and lowercase letters to designate matrices, column vectors and scalars, respectively. $\bm{X}^\dag$ denotes the complex conjugate transpose of matrix $\bm{X}$. $\bm{X}(:,i)$ denotes the $i$-th column of $\bm{X}$. $x_i$ denotes the $i$-th element of vector $\bm{x}$. $\textnormal{diag}[x_1,x_2,...,x_n]$ denotes a diagonal matrix with $x_1,x_2,...,x_n$ on its diagonal. $\det(\bm{X})$ and $\textnormal{tr}(\bm{X})$ denote the determinant and the trace of matrix $\bm{X}$, respectively. $\mathcal{CN}(\bm{\mu},\bm{\Sigma})$ denotes a circularly symmetric complex Gaussian random vector of mean $\bm{\mu}$ and covariance matrix $\bm{\Sigma}$. The logarithm $\log(x)$ denotes the binary logarithm. We use Cov($\cdot$) to denote the covariance matrix of a random vector.

\section{System Model}
\label{sec_SM}
We consider a downlink BC, where an $M$-antenna BS serves $N$ single-antenna users. The receive signal of the $n$-th user is expressed as
\begin{equation}
\label{channel_model}
y_n =  \bm{h}_n^\dag \bm{W} \bm{s} + n_n,
\end{equation}
where $\bm{s}\in\mathcal{C}^N$ is the data transmitted to the users, $\bm{x} = \bm{W}\bm{s}$ denotes the precoded downlink signals, $\bm{W}\in\mathcal{C}^{M \times N}$ denotes the precoding matrix, and $\bm{y}\in\mathcal{C}^{N}$ are the received signals of users. The downlink total transmit power constraint is
\begin{equation}
\label{Power_cons}
\textrm{tr}\left\{ \mathbb{E}[\bm{W}\bm{s} \bm{s}^\dag\bm{W}^\dag]\right\} \le P,
\end{equation}
and $\bm{n} \sim \mathcal{CN}(\bm{0},\bm{I}_N)$ is the Gaussian distributed uncorrelated noise.

\subsection{Spatial Correlated Channel Matrix}
Define the compound downlink channel matrix $\bm{H} = \left[\bm{h}_1,\bm{h}_2,...,\bm{h}_N\right]^\dag$, where $\bm{h}_n \sim \mathcal{CN}(\bm{0},\bm{R}_n)$. The CCM of user $n$ is
\begin{equation}
\label{channel_corr}
\bm{R}_n = \mathbb{E}\left[\bm{h}_n\bm{h}_n^\dag\right],
\end{equation}
where by the Karhunen-Loeve representation,
\begin{equation}
\label{KL}
\bm{h}_n = \bm{R}_n^{\frac{1}{2}}\bm{z}_n,
\end{equation}
where $\bm{z}_n \sim \mathcal{CN}(\bm{0},\bm{I}_M)$. It is assumed that the channel vectors of users are mutually \emph{independent}, since users are usually well separated. Denote the singular-value-decomposition (SVD) of the CCM as $\bm{R}_n = \bm{U}_n\bm{\Sigma}_n\bm{U}_n^\dag$, and $\bm{\Sigma}_n=\textrm{diag}[\lambda_1^{(n)}, \lambda_2^{(n)},..., \lambda_M^{(n)}]$. It is worthwhile to mention that in this work, we assume the BS and the users have \emph{perfect} knowledge of the second-order channel statistics, i.e., the CCMs.
\subsection{Dominant Eigenspace Representation of CCM}
Let us define the order-$r_n$ dominant eigenspace representation of $\bm{R}_n$ ($r_n$-DER) as
\begin{equation}
\label{channel_corr_eff}
\bm{R}_n^{(r_n)} = \bm{U}_n^{(r_n)}\bm{\Sigma}_n^{(r_n)}(\bm{U}_n^{(r_n)})^\dag,
\end{equation}
where $\bm{\Sigma}_n^{(r_n)} \in \mathcal{C}^{r_n \times r_n}$ contains the $r_n$ dominant eigenvalues, and  $\bm{U}_n^{(r_n)} \in \mathcal{C}^{M \times r_n}$ denotes the corresponding $r_n$ eigenvectors of $\bm{R}_n$. The order-$r_n$ channel vector approximation ($r_n$-CVA) is
\begin{equation}
\label{channel_eff_KL}
\bm{h}_n^{{(r_n)}} = \bm{U}_n^{{(r_n)}}(\bm{\Sigma}_n^{{(r_n)}})^{\frac{1}{2}}\bm{z}_n^{{(r_n)}},
\end{equation}
where $\bm{z}_n^{(r_n)} \sim \mathcal{CN}(\bm{0},\bm{I}_{r_n})$. And let
\begin{equation}
\label{channel_dr}
\bm{h}_n = \bm{h}_n^{{(r_n)}} +\bm{e}_n^{{(r_n)}},
\end{equation}
where $\bm{e}_n^{{(r_n)}}$ denotes the error introduced by only considering the dominant $r_n$ eigenvalues, which, therefore, can be represented as
\begin{equation}
\label{channel_eff_err}
\bm{e}_n^{{(r_n)}} =  \bm{\bar U}_n^{(r_n)}(\bm{\bar \Sigma}_n^{(r_n)})^{\frac{1}{2}}\bm{\bar z}_n^{(r_n)},
\end{equation}
where $\bm{\bar U}_n^{(r_n)} \in \mathcal{C}^{M \times (M - r_n)}$ denotes the remaining $M - r_n$ eigenvectors of $\bm{R}_n$, and $\bm{\bar \Sigma}_n^{(r_n)} \in \mathcal{C}^{(M - r_n) \times (M-r_n)}$ contains the remaining $M-r_n$ non-dominant eigenvalues. The approximation, namely the $r_n$-DER, which only accounts for the dominant $r_n$ eigenvalues of $\bm{R}_n$ is leveraged to improve the CSIT feedback efficiency, which is discussed in details in Section \ref{Sec_feedback}.
\section{FDD Massive MIMO Achievable Rates}
\label{sec_AE}
In this section, we will specify the rate-achieving transmission scheme proposed in this work. The structure of the transmission strategy is identical with the widely adopted pilot-assisted FDD multiuser-MIMO system, which encompasses three steps:
\begin{itemize}
\item
Downlink channel training.
\item
Uplink CSIT feedback.
\item
Data transmission.
\end{itemize}
The rate improvement stems from optimizing the channel training sequences and the CSIT feedback codebooks under the spatially correlated channels, thus requiring minimum modifications to current transmission strategy. In what follows, we will investigate the aforementioned steps in order, namely the channel training sequences, feedback codebooks, and derive the achievable rates on account of the dimensionality loss and imperfection of channel estimations with the RZF linear precoder.
\subsection{Optimal Downlink Training with Per-User CCM}
\label{sec_training}
The signal model of the channel training phase is expressed as
\begin{IEEEeqnarray}{rl}
\label{traning_sig}
& \bm{Y}_\tau = \bm{H} \bm{X}_\tau + \bm{N}_\tau \nonumber\\
& \textrm{tr}\left[\bm{X}_\tau \bm{X}_\tau^\dag \right] \le \tau P,
\end{IEEEeqnarray}
where $\bm{X}_\tau$ is a $M \times \tau$ training signal matrix, containing the training sequences and is known to the BS and the users. $\tau$ is the training length, and $\bm{Y}_\tau=\left[\bm{y}_{\tau,1},\bm{y}_{\tau,2},...,\bm{y}_{\tau,N}\right]^\dag$ is the corresponding channel output observed by the user, disturbed by Gaussian noise $\bm{N}_\tau$ with i.i.d. unit variance entries. The $n$-th user observes
\begin{equation}
\label{training_sig_n}
\bm{y}_{\tau,n}^{\dag} = \bm{h}_n^{\dag} \bm{X}_\tau + \bm{n}_{\tau,n}^\dag,
\end{equation}
and applies the minimum-mean-square-error (MMSE) estimation \cite[Section 19.5]{Molisch}
\begin{equation}
\label{training_output}
\bm{\hat h}_n^{\dag} = \bm{R}_n \bm{X}_\tau (\bm{X}_\tau^\dag \bm{R}_n \bm{X}_\tau + \bm{I}_\tau)^{-1} \bm{y}_{\tau,n}^{\dag}.
\end{equation}
Notice that we assume the CCMs are known to both the users and the BS. Applying the MMSE decomposition, the user channel $\bm{h}_n$ and the covariance matrix of the channel estimation error due to imperfect channel training are expressed as \cite{Caire10}
\begin{IEEEeqnarray}{rl}
\label{MMSE_error}
\bm{h}_n &= \bm{\hat h}_n + \bm{\hat{e}}_n, \nonumber \\
\bm{C}_{{\hat e}_n} &= (\bm{R}_n^{-1} + \bm{X}_\tau \bm{X}_\tau^\dag)^{-1},
\end{IEEEeqnarray}
respectively. The total mean-square error (MSE) is
\begin{equation}
\label{MSE}
\textrm{MSE} = \sum\limits_{n = 1}^N \textrm{tr}\left[{\bm{C}_{{{\hat e}_n}}}\right].
\end{equation}
Notice that by assumption $\bm{R}_n$ is the CCM, thus it may be rank-deficient and not invertible. Nonetheless, let $\bm{\bar R}_n = \bm{R}_n + \epsilon \bm{I}_M$ such that $\epsilon$ is small but $\bm{\bar R}_n$ is invertible. Then \eqref{MMSE_error} holds true if we substitute $\bm{\bar R}_n$ for $\bm{R}_n$. Then we can let $\epsilon \to 0$ due to the continuity of the function involved.

In \cite{Kotecha04}, the optimal training sequences where users have \emph{identical} CCMs are given, in the sense of minimizing the MSE or the mutual information between the channel coefficients and received signals conditioned on the transmitted block signals. However, to the best of our knowledge, the optimal training sequences under the per-user CCMs is still unknown, because multiple users share the same downlink training sequences, and thus the training sequence can no longer match one specific CCM, as in the case where user CCMs are identical \cite{Kotecha04}. In what follows, we develop an iterative algorithm to find the optimal training sequences, in terms of maximizing the conditional mutual information (CMI) between the channel vector and the received signal. The optimization problem, given the training length $\tau$ and total transmit power $P$ is first expressed as,
\begin{IEEEeqnarray}{rll}
\label{CMI}
&\textrm{maximize:\quad\quad} & \sum\limits_{n = 1}^N \log \det \left(\bm{I} + \bm{X}_\tau^\dag {\bm{R}_n}{\bm{X}_{\tau}}\right) \nonumber \\
&\textrm{s.t.} & \textrm{tr}\left[\bm{X}_\tau \bm{X}_\tau^\dag \right] \le \tau P,
\end{IEEEeqnarray}
and we have the following theorem.
\begin{theorem}
The training sequences that maximize the CMI satisfy following condition
\begin{equation}
\label{opt_ts}
\sum\limits_{n = 1}^N \left[\bm{R}_n \bm{X}_\textrm{opt}\left(\bm{I}_\tau+\bm{X}_\textrm{opt}^\dag \bm{R}_n \bm{X}_\textrm{opt}\right)^{-1}\right] = \lambda \bm{X}_\textrm{opt},
\end{equation}
where $\lambda \ge 0$ is a constant chosen to satisfy the power constraint.
\end{theorem}
\begin{IEEEproof}
The proof is straightforward by deriving the KKT conditions of the Lagrangian dual problem of \eqref{CMI}.
\end{IEEEproof}
\begin{remark}
Unfortunately, in general, the problem \eqref{CMI} is not a convex problem. Consider the special case where $N=1$ and $\bm{R}_1$ is rank-deficient, then any $\bm{X}_\tau$ satisfying
\begin{equation}
\label{badcase}
\bm{X}_\tau = \left[\bm{x}_0,\bm{x}_0,,...,\bm{x}_0\right],
\end{equation}
where $\bm{x}_0$ is the eigenvector of $\bm{R}_1$ corresponding to the eigenvalue of $0$, is the solution of \eqref{opt_ts} when $\lambda = 0$. Therefore, there are multiple sequences that satisfy the KKT condition in \eqref{opt_ts}, and clearly, none of which satisfying \eqref{badcase} is the optimal solution, since by plugging \eqref{badcase} into \eqref{CMI}, the objective is zero. To obtain an improved performance, we develop a heuristic iterative algorithm which is based on the condition in \eqref{opt_ts} to find the optimal training sequences, and based on the simulation results, the algorithm performs fairly well and converges fast.
\end{remark}
\begin{remark}
Observing the condition in \eqref{opt_ts}, one can immediately infer that when $N=1$, the optimal training sequences developed in \cite{Kotecha04} based on identical CCM, which contain the eigenvectors of the CCM with optimal power allocation given by the water-filling solution, satisfy \eqref{opt_ts}, i.e., the identical CCM is a special case for our problem.
\end{remark}
\begin{remark}
The reason that we set the objective to be maximizing the CMI, rather than directly minimizing the total MSE, is that the algorithm based on minimizing the MSE does \emph{not} converge. This non-convergent behavior is the result of the ill-conditioned matrices involved in computing the KKT conditions in the MSE problem. Consider the derivative of the MSE
\begin{equation}
\frac{{\partial \textrm{MSE}}}{{\partial \bm{X}_\tau}} = \sum\limits_{n = 1}^N \left(\bm{R}_n^{-1}+\bm{X}_\tau \bm{X}_\tau^\dag\right)^{-2} \bm{X}_\tau.
\end{equation}
The matrix $\left(\bm{R}_n^{-1}+\bm{X}_\tau \bm{X}_\tau^\dag\right)$ is often ill-conditioned, when $\bm{R}_n$ is rank-deficient, whereas in the CMI problem, the matrices involved are all well-conditioned. Moreover, based on \cite{Guo05}, the MMSE and the mutual information has very strong relationships, and the numerical results show that the obtained training sequences have very good MSE performance.
\end{remark}
The iterative algorithm, which aims to find the optimum training sequences based on the first-order KKT condition in \eqref{opt_ts} is specified as follows
\begin{itemize}
\item
Step 1) Initialization:
\begin{IEEEeqnarray}{rl}
\label{initial}
\bm{X}_1=\bm{X}_0;
\end{IEEEeqnarray}
\item
Step 2) Iteration:
\begin{equation}
\bm{X}_i = \sum\limits_{n = 1}^N \left[\bm{R}_n \bm{X}_{i-1}\left(\bm{I}_\tau+\bm{X}_{i-1}^\dag \bm{R}_n \bm{X}_{i-1}\right)^{-1}\right].
\end{equation}
Then apply the power normalization, where
\begin{equation}
\bm{X}_i \leftarrow  \frac{\tau P}{\textrm{tr}\left[\bm{X}_i\bm{X}_i^\dag\right]}\bm{X}_i.
\end{equation}
If $\|\bm{X}_i-\bm{X}_{i-1}\|<\epsilon$, the algorithm is finished, and the output is the training sequences $\bm{X}_i$. Else, go to step $2$.
\end{itemize}
\begin{remark}
Notice that $\bm{X}_0 \neq 0$, otherwise the algorithm would be stuck at zero. In our simulations, letting $\bm{X}_0$ orthogonal rows works well. Also notice that in the algorithm, we normalize the power of the training signals to be equal to the power constraint, due to the fact that it is clear that the optimal solution satisfies the power constraint with equality.
\end{remark}

For comparison purposes, the simulations also consider the unitary training sequences, which are shown to be optimal with i.i.d. channels \cite{Hassibi03}. We assume\footnote{Notice that the unitary training sequences for the case $\tau<M$ is not well defined in \cite{Hassibi03}, since it does not suffice to have $\tau<M$ pilots in i.i.d. channels. Here we assume $X_\tau$ has orthogonal rows when $\tau<M$.}
\begin{equation}
\left\{ \,
\begin{IEEEeqnarraybox}[][c]{l?s}
\IEEEstrut
\bm{X}_\tau \bm{X}_\tau^\dag = \frac{\tau P}{M}\bm{I}_M & if $\tau \ge M$, \\
\bm{X}_\tau^\dag \bm{X}_\tau = P\bm{I}_\tau & if $\tau < M$,
\IEEEstrut
\end{IEEEeqnarraybox}
\right.
\end{equation}
The comparison is shown in Section \ref{sec_nr}.
\subsection{Uplink CSIT Feedback}
\label{Sec_feedback}
After the users estimate their respective channel coefficients based on received channel training signals, they feed back their estimates using predefined codebooks. In this subsection, efficient channel feedback schemes are exploited with spatially correlated channels. We first propose the entropy encoded scalar quantization after KL transform, which is a simple way to universally approach the optimal VQ performance. Then we compare its performance with two VQ approaches, which are shown to be near-optimal with spatially correlated channels and also serve as two implementation options.
\subsubsection{Entropy Coded Scalar Quantization}
We consider a scalar quantization (component by component) of the transformed channel vector. Specifically, denote
\begin{equation}
\bm{\hat h}_n^{\textrm{KL}} = \bm{U}_n^\dag \bm{\hat h}_n = \bm{\Sigma}^{\frac{1}{2}}_n \bm{z}_n - \bm{U}_n^\dag \bm{\hat e}_n
\end{equation}
as the KL-transform of the channel vector of user-$n$, after channel traning. Putting aside the channel training error $\bm{\hat e}_n$, this yields $M$ mutually independent Gaussian variables with non-identical variances. The reverse water-filling approach (RWF) \cite{Cover12} can be implemented to achieve the rate-distortion function in this scenario, i.e., we allocate the quantization bits according to the following conditions
\begin{IEEEeqnarray}{rcl}
\label{RWF}
&& \sum_{i=1}^M {\min{\left[\gamma,\lambda_i^{(n)}\right]}} = D \nonumber \\
&& R_i = \log\left(\frac{\lambda_i^{(n)}}{\gamma}\right) \nonumber \\
&& \sum_{i=1}^M {R_i} = B_n,
\end{IEEEeqnarray}
where $D$ is the total MSE distortion, $R_i$ denotes the number of bits allocated to the $i$-th component of $\bm{\hat h}_n^{\textrm{KL}}$, $B_n$ is the total number of feedback bits for user-$n$, and $\gamma$ denotes the water level. The MSE distortion for the $i$-th component is
\begin{equation}
D_i = \min{\left[\gamma,\lambda_i^{(n)}\right]}.
\end{equation}
After the BS recovers the KL-transformed channel vector from the user feedback, it can reconstruct the channel vector by the inverse KL-transform. By this scheme, we obtain the relationship between the channel estimation at the BS side and the real channel, i.e.,
\begin{IEEEeqnarray}{rcl}
\bm{h_n} &=& \bm{\hat {\bar h}}_{n} + \underbrace{\bm{\hat e}_n + \bm{U}_n \bm{\hat{\bar e}}_n}_{\bm{\varepsilon}_n}, \\
\label{SQ_er}
\textrm{Cov}(\bm{\varepsilon}_n) &=&  \underbrace{\bm{C}_{{\hat e}_n} }_{\mathcal{M}_1} + \underbrace{\bm{U}_n \bm{D}_n\bm{U}_n^\dag}_{\mathcal{M}_2}  \\
\bm{{\hat{\bar R}}}_n &\triangleq& \textrm{Cov}(\bm{\hat {\bar h}}_{n}) = \bm{R}_n-\textrm{Cov}(\bm{\varepsilon}_n),
\end{IEEEeqnarray}
where $\bm{C}_{\hat e_n}$ is defined in \eqref{MMSE_error}, and $\bm{D}_n \triangleq \textrm{diag}\left[D_1, D_2,..., D_M\right]$. Observing the error covariance matrix in \eqref{SQ_er}, $\mathcal{M}_1$ and $\mathcal{M}_2$ represent channel estimation error due to imperfect channel training and CSI quantization error respectively.
\begin{remark}
There are several approaches to mimic such behavior using a scalar quantizer, e.g., apply a Huffman code on each of the components with $\lambda_i^{(n)}>\gamma$, based on the fact that the component is Gaussian distributed with variance $\lambda_i^{(n)}$. The advantage of this quantizer is that it does not involve any VQ, thus can be implemented very efficiently in parallel. Notice also that when $\bm{U}_n$ is a slice of a DFT matrix (as in large linear antenna arrays), the KL-transform can be well approximated by an FFT, therefore the overall quantization can be made extremely computationally efficient. The MSE performance and comparison with VQ approaches will be shown in Section \ref{sec_TF}.
\end{remark}
\subsubsection{VQ: Isotropical and Skewed Random Codebooks}
In the literature, extensive work has been done regarding the VQ feedback codebook design in spatial CCMs. It is well understood that in the asymptotic regime where the number of feedback bits $B$ goes to infinity, the MSE scales down with $B$ as MSE $\sim 2^{\frac{-B}{M-1}}$, regardless whether the channel distribution is i.i.d. or correlated \cite{Raghavan13}\cite{Chun07}. However, when the number of feedback bits $B$ is limited, which is the case for FDD massive MIMO systems due to scarce channel estimation resources, the exact analysis for the MSE performance is unavailable. In \cite{Xia06}, a ``skewed codebook'' (i.e., a codebook based on skewing an isotropical codebook) that matches the eigenspace of the CCM, is shown to be close to optimal by simulation results. The authors of \cite{Raghavan13} try to derive closed-form expressions for the SNR loss for general skewed codebooks, but the expressions are too complicated to find the optimal skew matrix in closed form. Notwithstanding the difficulty in deriving the optimal codebook in closed form, the Lloyd algorithm can be implemented to find the optimal codebook, however with high computational complexity \cite{Lo99}.

Observing that the CCMs of the users are usually rank-deficient, in the sense that a number of eigenvalues of the CCMs are extremely small (see numerical results in Section \ref{sec_nr} for eigenvalue distributions in popular channel models), it is advantageous for the users to compress their feedback overhead by \emph{only} feeding back along the order-$r_n$ dominant eigenspace of the channel, i.e., a $r_n$-CVA in \eqref{channel_eff_KL}. It will be shown later that this scheme performs better than feeding back all the channel space, when $B$ is finite. Specifically, we consider two kinds of feedback schemes, both of which concentrate the feedback bits in the dominant eigenspace of the channels, while one of them leverages an isotropical random vector to quantize the dominant eigenspace, the other explores the benefit of a skewed codebook design.

\paragraph{Isotropical Quantization in Dominant Eigenspace}
First, the $n$-th user decorrelates the channel vector leveraging the $r_n$-DER of the CCM,
\begin{equation}
\label{z_hat_est}
\bm{\hat z}_{n}^{(r_n)} =  (\bm{\Sigma}_n^{{(r_n)}})^{-\frac{1}{2}}(\bm{U}_n^{{(r_n)}})^\dag\bm{\hat h}_n.
\end{equation}
Notice that assuming the $r_n$-CVA is accurate and the channel training is perfect, namely $\bm{\hat h}_n = \bm{h}_n^{{(r_n)}}$, then $\bm{\hat z}_{n}^{(r_n)}$ has $r_n$ independently Gaussian distributed unit-norm entries. Based on this observation, we then use a predefined isotropical codebook to quantize $\bm{\hat z}_{n}^{(r_n)}$. After the feedback, the BS obtains a quantized version of the channel estimation, after multiplying the channel correlation eigenvectors,
\begin{equation}
\label{h_hat_BSest1}
\bm{\hat {\bar h}}_{n} =  \bm{U}_n^{{(r_n)}}(\bm{\Sigma}_n^{{(r_n)}})^{\frac{1}{2}}\bm{\hat{ \bar z}}_n^{(r_n)},
\end{equation}
where $\bm{\hat{ \bar z}}_n^{(r_n)}$ denotes the quantized version of $\bm{\hat{ z}}_n^{(r_n)}$ at the BS side, with quantization error $\bm{\hat{\bar e}}_n$ satisfying
\begin{equation}
\label{h_quant}
\bm{\hat{ z}}_n^{(r_n)} = \bm{\hat{\bar z}}^{(r_n)}_n + \bm{\hat{\bar e}}_n.
\end{equation}
The quantization error $\bm{\hat{\bar e}}_n$ can be computed based on \cite{Chun07}, where random vector quantization (RVQ) is assumed, by which the codebook is obtained by generating $2^{B_n}$ quantization vectors independently and uniformly distributed on the unit sphere in $\mathcal{C}^{r_n}$. The quantization error $\bm{\hat{\bar e}}_n$ is i.i.d. and independent with $\bm{\hat{\bar z}}^{(r_n)}_n$. It follows that
\begin{equation}
\label{h_quant_error_cov}
\textrm{Cov}({\bm{\hat{\bar e}}_n}) = \frac{2^{\frac{-B_n}{r_n-1}}}{r_n} \beta \bm{I}_{r_n},
\end{equation}
where
\begin{equation}
\label{h_quant_error}
\beta = \textrm{tr} \left[\bm{\hat{ z}}_n^{(r_n)}(\bm{\hat{ z}}_n^{(r_n)})^\dag\right] = \textrm{tr} \left[\bm{I}_{r_n}-(\bm{\Sigma}_n^{{(r_n)}})^{-\frac{1}{2}}(\bm{U}_n^{{(r_n)}})^\dag\bm{C}_{{\hat e}_n}\bm{U}_n^{{(r_n)}}(\bm{\Sigma}_n^{{(r_n)}})^{-\frac{1}{2}}\right].
\end{equation}
Combining \eqref{MMSE_error}, \eqref{z_hat_est}, \eqref{h_hat_BSest1}, \eqref{h_quant_error_cov} and the $r_n$-CVA in \eqref{channel_dr}, we obtain the relationship between the channel estimation at the BS side and the real channel, i.e.,
\begin{IEEEeqnarray}{rcl}
\bm{h_n} &=& \bm{\hat {\bar h}}_{n} + \underbrace{\bm{U}_n^{{(r_n)}}(\bm{U}_n^{{(r_n)}})^\dag \bm{\hat e}_n + \bm{U}_n^{{(r_n)}} (\bm{\Sigma}_n^{{(r_n)}})^{\frac{1}{2}} \bm{\hat{\bar e}}_n+\bm{e}_n^{(r_n)}}_{\bm{\varepsilon}_n}, \\
\label{h_hat_BSest2}
\textrm{Cov}(\bm{\varepsilon}_n) &=&  \underbrace{\bm{U}_n^{{(r_n)}}(\bm{U}_n^{{(r_n)}})^\dag\bm{C}_{{\hat e}_n}\bm{U}_n^{{(r_n)}}(\bm{U}_n^{{(r_n)}})^\dag}_{\mathcal{M}_1} + \underbrace{ \frac{2^{\frac{-B_n}{r_n-1}}}{r_n} \beta \bm{R}_n^{(r_n)}}_{\mathcal{M}_2} + \underbrace{\bm{\bar U}_n \bm{\bar \Sigma}_n \bm{\bar U}_n^\dag}_{\mathcal{M}_3} \\
\bm{{\hat{\bar R}}}_n &\triangleq& \textrm{Cov}(\bm{\hat {\bar h}}_{n}) = \bm{R}_n-\textrm{Cov}(\bm{\varepsilon}_n),
\end{IEEEeqnarray}
where $\bm{C}_{\hat e_n}$ is defined in \eqref{MMSE_error}. Observing the error covariance matrix in \eqref{h_hat_BSest2}, $\mathcal{M}_1$, $\mathcal{M}_2$, $\mathcal{M}_3$ represent channel estimation error due to imperfect channel training, CSI quantization error, and the error from only feeding back the order-$r_n$ dominant eigenspace of the channel vectors, respectively.
\paragraph{Skewed Codebook in Dominant Eigenspace}
Although we concentrate our feedback bits in the dominant eigenspace based on the isotropical dominant codebook design in the preceding subsection, there is still \emph{imbalance} among the eigenvalues of the CCMs, rendering the isotropical RVQ codebook described above not optimal. To this end, we adopt a skewed codebook
\begin{equation}
\label{sk_codebook}
\mathfrak{C}_{\textrm{sk}}=\left\{\frac{\bm{A}^\frac{1}{2}_n \bm{f}_i}{\left\| \bm{A}^\frac{1}{2}_n \bm{f}_i \right\|}, \,  i=1,...,2^B\right\}£¬
\end{equation}
where $\bm{f}_i \in \mathcal{C}^{r_n}$ is isotropically distributed on the unit-sphere, and $\bm{A}_n = \bm{U}_n^{{(r_n)}}(\bm{\Sigma}_n^{{(r_n)}})^{\frac{1}{2}}$. It is clear that by design we only feed back the dominant $r_n$ eigenmodes of the channel, i.e., $\bm{h}_n^{{(r_n)}}$, neglecting the remaining eigenmodes.  The skewed matrix is designed to match the dominant eigenspace of the channel, such that the correlation matrix of the codebook is identical with the $r_n$-DER. By adopting the codebook design, the total \emph{quantization error}, which is defined as
\begin{equation}
\textrm{MSE}_\textrm{q} = \textrm{tr}\left[\mathbb{E}\left(\bm{\hat{\bar e}}_n^\dag \bm{\hat{\bar e}}_n\right)\right],
\end{equation}
can be upper bounded based on the following theorem.
\begin{theorem}
\label{thm_mseq}
Given a channel vector $\bm{h}_n$, the quantization error based on the skewed codebook defined in \eqref{sk_codebook} is upper bounded as
\begin{equation}
\label{MSE_QSK}
\textrm{MSE}_\textrm{q} \le \frac{{\sum\limits_{i = 1}^{{r_n}} {(\lambda _i^{(n)})^2} }}{{{\lambda _1^{(n)}}}}{2^{\frac{{ - {B_n}}}{{{r_n} - 1}}}} + \textrm{tr} \left[ \bm{\bar U}_n \bm{\bar \Sigma}_n \bm{\bar U}_n^\dag \right].
\end{equation}
\end{theorem}
\begin{IEEEproof}
The proof is based upon the distribution results developed for the i.i.d. channels in \cite{Chun07}. The detail proof is in Appendix \ref{App_thm_mseq}.
\end{IEEEproof}
\begin{remark}
It is clear that the first and second terms in \eqref{MSE_QSK} represent the error resulting from quantizing the channel and neglecting the subdominant eigenmodes of the channel, respectively. Also notice that the quantization error by the skewed codebook is smaller than the MSE by isotropical codebook,
\begin{equation}
\textrm{MSE}_\textrm{q,sk} \le \sum\limits_{i = 1}^{{r_n}} {\lambda _i^{(n)}} {2^{\frac{{ - {B_n}}}{{{r_n} - 1}}}} + \textrm{tr} \left[ \bm{\bar U}_n \bm{\bar \Sigma}_n \bm{\bar U}_n^\dag \right] = \textrm{MSE}_\textrm{q,iid},
\end{equation}
where the equality holds if and only if $\lambda_1^{(n)} = \lambda_2^{(n)} =...=\lambda_{r_n}^{(n)}$.
\end{remark}
\begin{remark}
Notice that the quantization error in Theorem \ref{thm_mseq} does not scale with $B$ to zero. This can be explained that when $B$ is large, it is better to quantize all the channel eigenmodes instead of neglecting the subdominant modes, i.e., $r_n=M$. Thus the quantization error with the optimal $r_n$, which minimizes the quantization error, scales with $B$ to zero, when $B$ goes to infinity. Meanwhile, the bound in Theorem \ref{thm_mseq} is tighter than the one with $r_n$ fixed to be $M$, when $B$ is finite. The numerical results in Section \ref{sec_nr} agrees with our analysis.
\end{remark}
\begin{remark}
Notice that the dominant rank $r_n$, i.e., the order of the CVA we choose to approximate the correlated channels, plays an important role in the feedback scheme. The larger $r_n$ is, the more accuracy we obtain by approximating the correlated channels, whereas the feedback quantization error is also larger due to the increased quantization dimension. Therefore, there exists a tradeoff in terms of the dominant rank, $r_n$. The optimal $r_n$ can be determined by a simple one-dimensional search over $1$:$M$, performed by the $n$-th user.
\end{remark}
\subsection{Data Transmission}
For fair comparison, also in line with the work in \cite{Adhikary13} and \cite{Wagner12}, we consider the RZF linear precoder schemes. The precoder treats the channel estimates as the real channel coefficients. Corresponding achievable rates on account of the imperfect channel estimations are computed in the following section. The RZF precoding matrix is expressed as
\begin{equation}
\label{RZF}
\bm{W}_{\textrm{rzf}} = \zeta \left(\bm{\hat {\bar H}}^\dag \bm{\hat {\bar H}}+M\alpha \bm{I}_M\right)^{-1}\bm{\hat {\bar H}}^\dag,
\end{equation}
where $\bm{\hat {\bar H}} = \left[\bm{\hat {\bar h}}_{1},\bm{\hat {\bar h}}_{2},...,\bm{\hat {\bar h}}_{N}\right]^\dag$, $\zeta$ is a normalization scalar to fulfill the power constraint in \eqref{Power_cons}, and $\alpha$ is the regularization factor. Based on \eqref{Power_cons}, we obtain
\begin{equation}
\label{zeta}
\zeta^2 = \frac{N}{\textrm{tr}\left[{\bm{\hat {\bar H}}{{\left( {{{\bm{\hat {\bar H}}}^\dag }\bm{\hat {\bar H}} + M\alpha {\bm{I}_M}} \right)}^{ - 2}}{{\bm{\hat {\bar H}}}^\dag }}\right]},
\end{equation}
where equal power allocation is assumed, i.e., $\left[\mathbb{E}[\bm{s} \bm{s}^\dag]\right]_{i,i}=\frac{P}{N}$. Define
\begin{equation}
\label{K}
\bm{K}_{\textrm{rzf}} = \left(\bm{\hat {\bar H}}^\dag\bm{\hat {\bar H}}+M\alpha \bm{I}_M\right)^{-1},
\end{equation}
the signal-to-interference-and-noise-ratio (SINR) of user $n$ is
\begin{equation}
\label{sinr_rzf}
{\gamma _{n,\textrm{rzf}}} = \frac{{{{\left| {\bm{\hat{\bar h}}_n^\dag {\bm{K}_\textrm{rzf}}{{\bm{\hat {\bar h}}}_n}} \right|}^2}}}{{\frac{N}{{P{\zeta ^2}}} + \left| {\bm{\varepsilon}_n^\dag {\bm{K}_\textrm{rzf}}{{\bm{\hat {\bar h}}}_n}} \right|^2 + \bm{h}_n^\dag {\bm{K}_{\textrm{rzf}}}\bm{\hat {\bar H}}_{[n]}^\dag {{\bm{\hat {\bar H}}}_{[n]}}{\bm{K}_{\textrm{rzf}}}{\bm{h}_n}}},
\end{equation}
where $\bm{\hat {\bar H}}_{[n]} = \left[\bm{\hat {\bar h}}_{1},...,\bm{\hat {\bar h}}_{n-1},\bm{\hat {\bar h}}_{n+1},...,\bm{\hat {\bar h}}_{N}\right]^\dag$. The training dimensionality loss is the length of the training sequence $\tau$. Assuming the feedback information is transmitted over the uplink MIMO-multiple-access-channel (MIMO-MAC), and based on \cite{Caire10}, the total feedback dimensionality loss is computed as
\begin{equation}
\label{feedback_oh}
\delta  = \frac{{\sum\limits_{n = 1}^N {{B_n}} }}{{{C_{\textrm{MIMO-MAC}}}}}.
\end{equation}
For the ease of exposition, we assume $B_n=B$, $\forall n$, and
\begin{equation}
C_{\textrm{MIMO-MAC}} = \kappa \min{\left[M,N\right]} \log(M\textsf{SNR}_\textrm{ul}),
\end{equation}
where $\kappa \in (0,1)$ is a scalar representing the diversity-multiplexing tradeoff in MIMO-MAC as defined in \cite{Caire10}.
The achievable sum rate considering imperfect channel training and feedback, $\bar{R}_{\textrm{rzf}}$, is expressed as the solution of the following optimization problem
\begin{IEEEeqnarray}{rll}
\label{AR}
& \textrm{maximize:\quad\quad} & \left(1-\frac{\tau+\delta}{T}\right)\sum\limits_{n = 1}^N {\log (1 + {\gamma _{n,\textrm{rzf}}})}  \nonumber \\
& \textrm{s.t.} & 1\le \tau+\delta \le T, \nonumber\\
&& \tau \ge 1,\quad \delta \ge 1,
\end{IEEEeqnarray}
where the optimization is over the training and feedback length. The fundamental tradeoff is larger training and feedback length provides a more accurate channel estimation whereas resulting in larger dimensionality loss. Since our focus is on the performance of the downlink BC achievable rates with correlated channels, we use an exhaustive search to find the optimal training and feedback length. The analysis of the optimal training and feedback length for i.i.d. channels can be found in \cite{Wagner12} and \cite{Kobayashi11}.
\section{Performance Analysis}
\label{sec_DE}
In this section, we provide expressions for the downlink achievable sum rate under the pre-user CCMs, leveraging the deterministic equivalent techniques provided in \cite{Wagner12}, with necessary modifications. For ease of exposition, we assume the dominant ranks we choose in the feedback schemes are identical, i.e., $r_n=r$, $\forall n$.

Following the approach in \cite{Wagner12}, when $M$ goes to infinity, the SINR of user $n$, $\gamma _{n,\textrm{rzf}}$, satisfies
\begin{equation}
\gamma _{n,\textrm{rzf}} - \gamma _{n,\textrm{rzf}}^o \stackrel{M \to \infty}{\longrightarrow} 0 \textrm{  with probability } 1,
\end{equation}
where $\gamma _{n,\textrm{rzf}}^o$ is a deterministic quantity that can be computed as
\begin{equation}
\label{DE}
\gamma _{n,{\rm{rzf}}}^o = \frac{{\frac{{{(\hat {\bar e}_n^o)^2}}}{{{{(1 + \hat {\bar e}_n^o)}^2}}}}}{\frac{{{\phi ^o}}}{P} + {{\hat {\bar E}_n^o} + {I_n^o}}},
\end{equation}
where the parameters involved are specified in \eqref{DE_w}-\eqref{DE_wend}. The derivation is mostly based upon \cite{Wagner12}, with generalizations to uncorrelated channel estimation error matrices. The details are omitted for brevity.
\newcounter{MYtempeqncnt}
\begin{figure*}[!t]
% ensure that we have normalsize text
\normalsize
% Store the current equation number.
\setcounter{MYtempeqncnt}{\value{equation}}
% Set the equation number to one less than the one
% desired for the first equation here.
% The value here will have to changed if equations
% are added or removed prior to the place these
% equations are referenced in the main text.
\begin{IEEEeqnarray}{lcl}
\label{DE_w}
{\phi ^o} &=& \frac{1}{M}\sum\limits_{n = 1}^N {\frac{{\hat {\bar e}{{_n^o}^\prime }}}{{{{(1 + \hat {\bar e}_n^o)}^2}}}},  \\
\hat {\bar e}_n^o &=& \frac{1}{M}\textrm{tr}\left[{\bm{{\hat {\bar R}}}_n}\bm{T}\right],  \\
\bm{T} &=& {\left(\frac{1}{M}\sum\limits_{j = 1}^N {\frac{{{\bm{{\hat {\bar R}}}_j}}}{{1 + \hat {\bar e}_j^o}}}  + \alpha {\bm{I}_M}\right)^{ - 1}},  \\
{\bm{{\hat {\bar e}}}^o}^\prime  &=& {\left[{\hat {\bar e}_1^o}^\prime ,{\hat {\bar e}_2^o}^\prime ,...,{\hat {\bar e}_N^o}^\prime \right]^T}= {\left({\bm{I}_N} - {\bm{J}}\right)^{ - 1}}{\bm{v}},  \\
{[\bm{J}]_{i,j}} &=& \frac{1}{M}\frac{{\frac{1}{M}\textrm{tr}\left[{\bm{{\hat {\bar R}}}_i}\bm{T}{\bm{{\hat {\bar R}}}_j}\bm{T}\right]}}{{{{\left(1 + \hat {\bar e}_j^o\right)}^2}}}, \\
\bm{v} &=& \frac{1}{M}{\left[\textrm{tr}\left({\bm{{\hat {\bar R}}}_1}{\bm{T}^2}\right),\textrm{tr}\left({\bm{{\hat {\bar R}}}_2}{\bm{T}^2}\right),...,\textrm{tr}\left({\bm{{\hat {\bar R}}}_N}{\bm{T}^2}\right)\right]^\textrm{T}} \\
{{\hat {\bar E}}_n^o} &=& \frac{{d_{n,n}^o}}{{M{{(1 + \hat {\bar e}_n^o)}^2}}},  \\
\bm{d}_n^o &=& \left[d_{n,1}^o,d_{n,2}^o,...,d_{n,N}^o\right]^T={\left({\bm{I}_N} - \bm{J}\right)^{ - 1}}\bm{b}_n,  \\
\bm{b}_n &=& \frac{1}{M}{\left[\textrm{tr}\left({\bm{{\hat {\bar R}}}_1}\bm{T}\left({\bm{R}_n} - {\bm{{\hat {\bar R}}}_n}\right)\bm{T}\right),...,\textrm{tr}\left({\bm{{\hat {\bar R}}}_N}\bm{T}\left({\bm{R}_n} - {\bm{{\hat {\bar R}}}_n}\right)\bm{T}\right)\right]^\textrm{T}},  \\
I_n^o &=& \frac{{{u_n}}}{{{{\left(1 + \hat {\bar e}_n^o\right)}^2}}} + \sum\limits_{j \ne n}^N {\frac{{d_{n,j}^o}}{{M{{\left(1 + \hat {\bar e}_j^o\right)}^2}}}}, \\
{u_n} &=& \frac{1}{M}\sum\limits_{j \ne n}^N {\frac{{f_{n,j}^o}}{{\left(1 + \hat {\bar e}_j^o\right)^2}}},\\
\bm{f}_n^o &=& \left[f_{n,1}^o,...,f_{n,N}^o\right]^T={\left({\bm{I}_N} - \bm{J}\right)^{ - 1}}\bm{c}_n,  \\
\bm{c}_n &=& \frac{1}{M}{\left[\textrm{tr}\left({\bm{{\hat {\bar R}}}_1}\bm{T}{\bm{{\hat {\bar R}}}_n}\bm{T}\right),...,\textrm{tr}\left({\bm{{\hat {\bar R}}}_N}\bm{T}{\bm{{\hat {\bar R}}}_n}\bm{T}\right)\right]^\textrm{T}}.
\label{DE_wend}
\end{IEEEeqnarray}
% Restore the current equation number.
\setcounter{equation}{55}
% IEEE uses as a separator
\hrulefill
% The spacer can be tweaked to stop underfull vboxes.
\vspace*{4pt}
\end{figure*}
\begin{figure}[!t]
\centering
\includegraphics[width=0.5\textwidth]{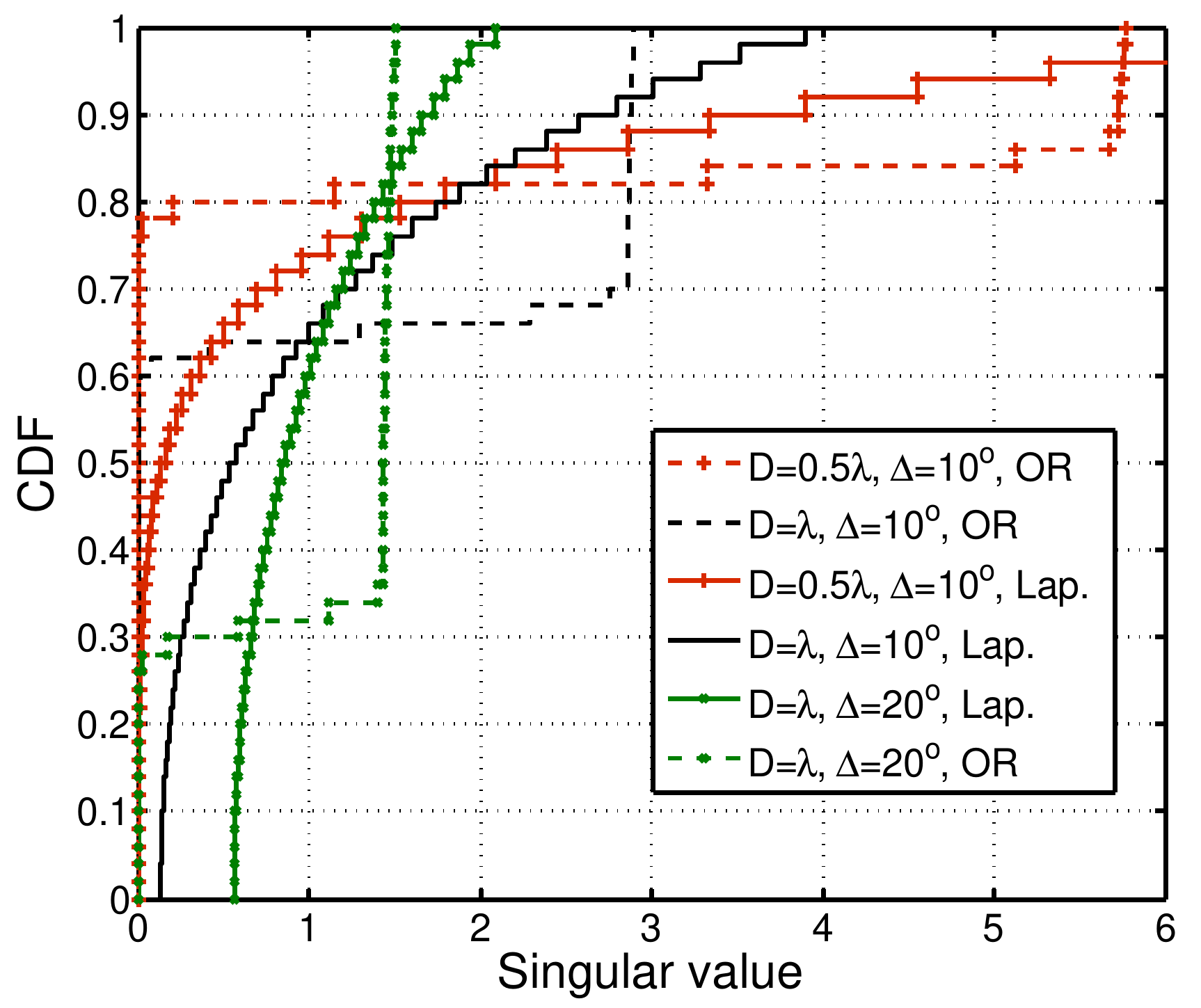}
\caption{The CDF of the singular values of user CCMs for various parameters. The number of BS antennas is $M=50$.}
\label{Fig_ER}
\end{figure}
\begin{figure}[!t]
\centering
\includegraphics[width=0.5\textwidth]{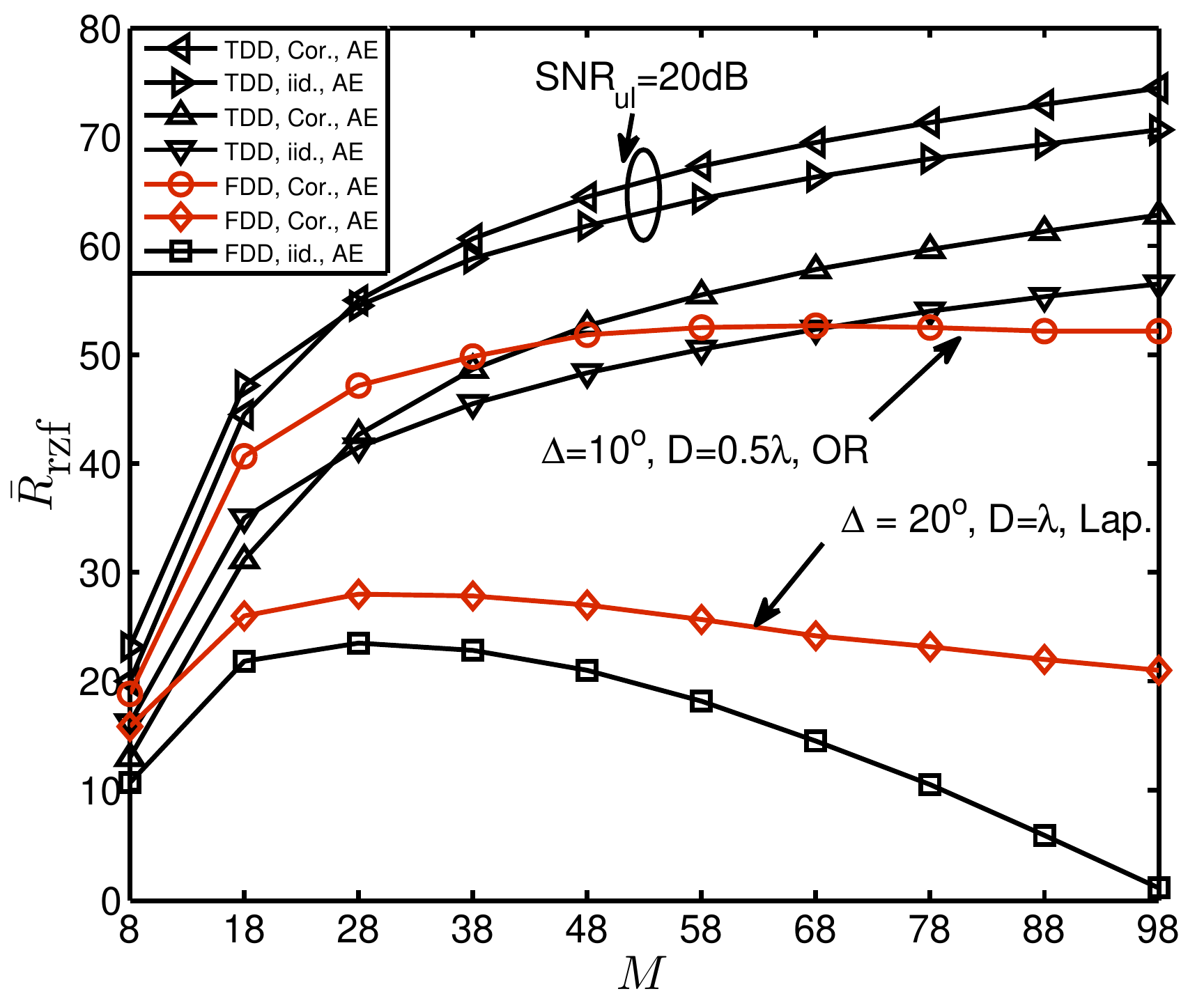}
\caption{Achievable sum rates in massive MIMO systems with i.i.d. channels, per-user correlation channels, TDD mode and FDD mode respectively. The downlink and uplink SNR are set to $20$~dB and $10$~dB, respectively, unless labeled otherwise. The channel block length is $T=200$. The number of users in the cell is $N=8$. The regularization factor of the RZF precoder is $\alpha = 0.01$. The per-user channel correlation matrices are calculated according to \eqref{OR} and \eqref{Lap}.}
\label{Fig_TDDvsFDD}
\end{figure}
\begin{figure}[!t]
\centering
\includegraphics[width=0.5\textwidth]{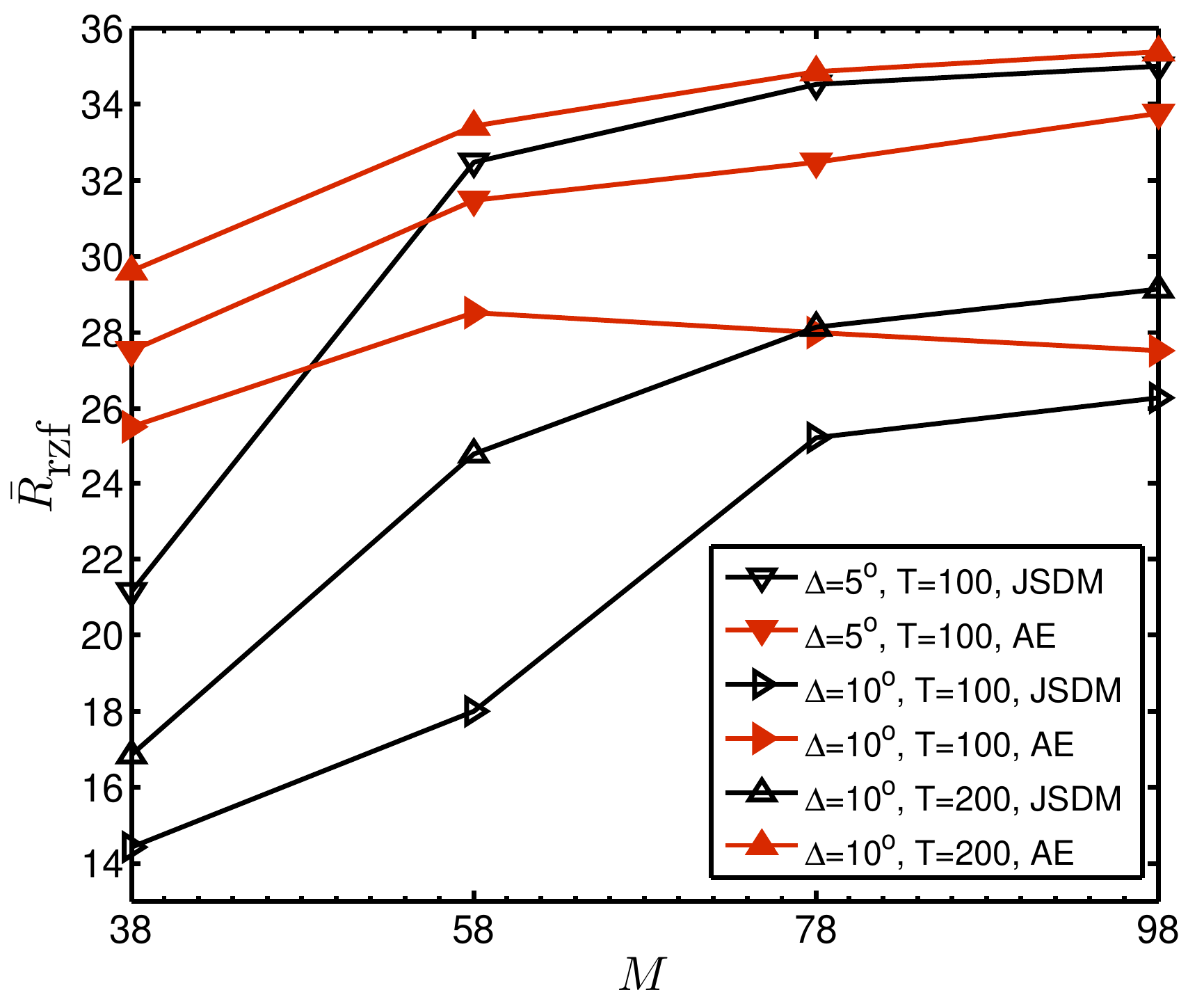}
\caption{The achievable sum rates (AE) obtained by the eigenspace channel estimation, compared to the JSDM scheme. The downlink and uplink SNR are both $10$~dB. The number of simultaneous users is $8$. }
\label{Fig_JSDM}
\end{figure}
\begin{figure}[!t]
\centering
\includegraphics[width=0.5\textwidth]{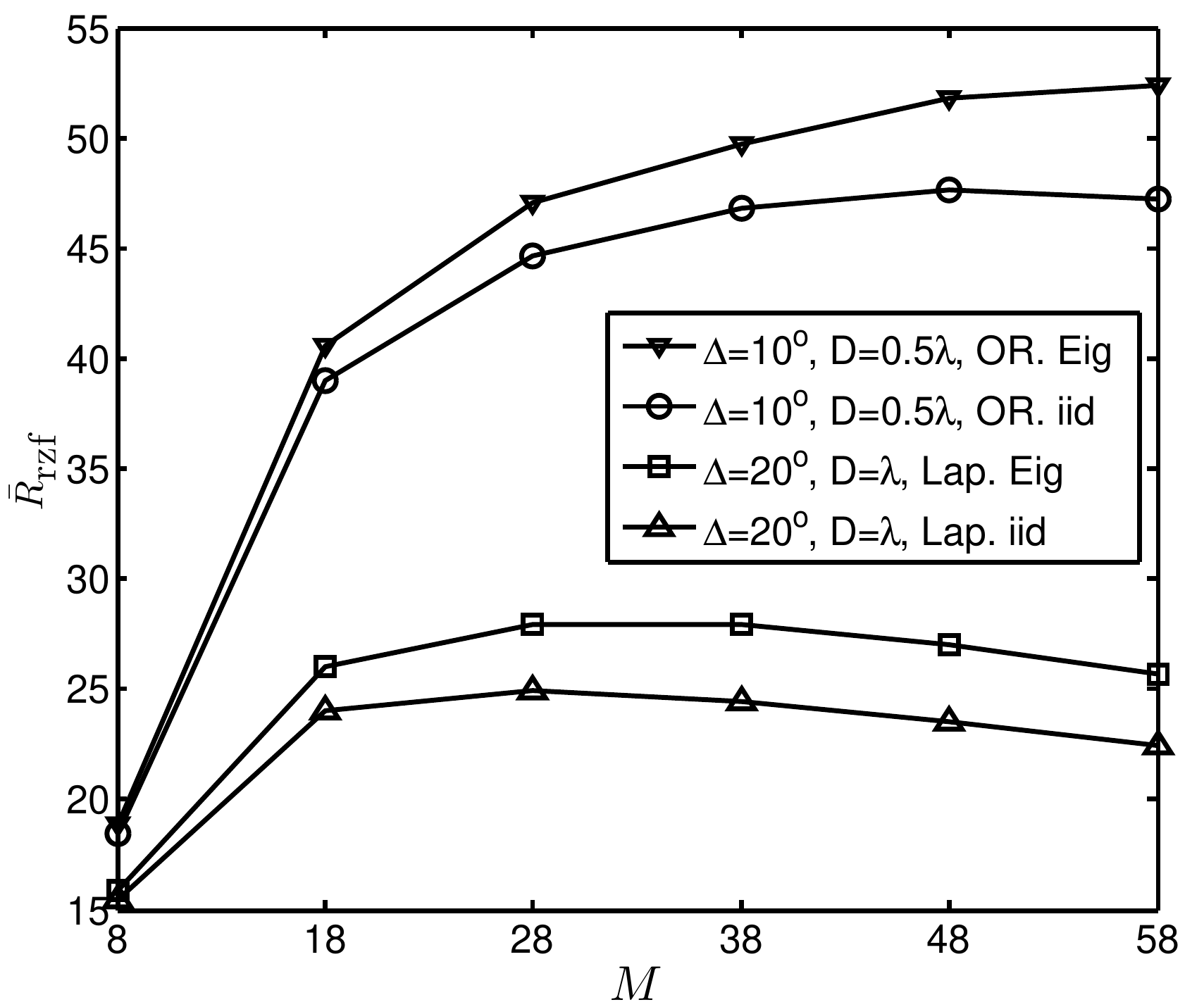}
\caption{Achievable sum rates in massive MIMO systems with eigenspace training and feedback schemes, compared with unitary channel estimation schemes commonly used for the i.i.d. channels. The downlink and uplink SNR are set to $20$~dB and $10$~dB, respectively. The channel block length is $T=200$. The number of users in the cell is $N=8$. The regularization factor of RZF precoder is $\alpha = 0.01$.}
\label{Fig_OPTvsORG}
\end{figure}
\begin{figure}[!t]
\centering
\includegraphics[width=0.5\textwidth]{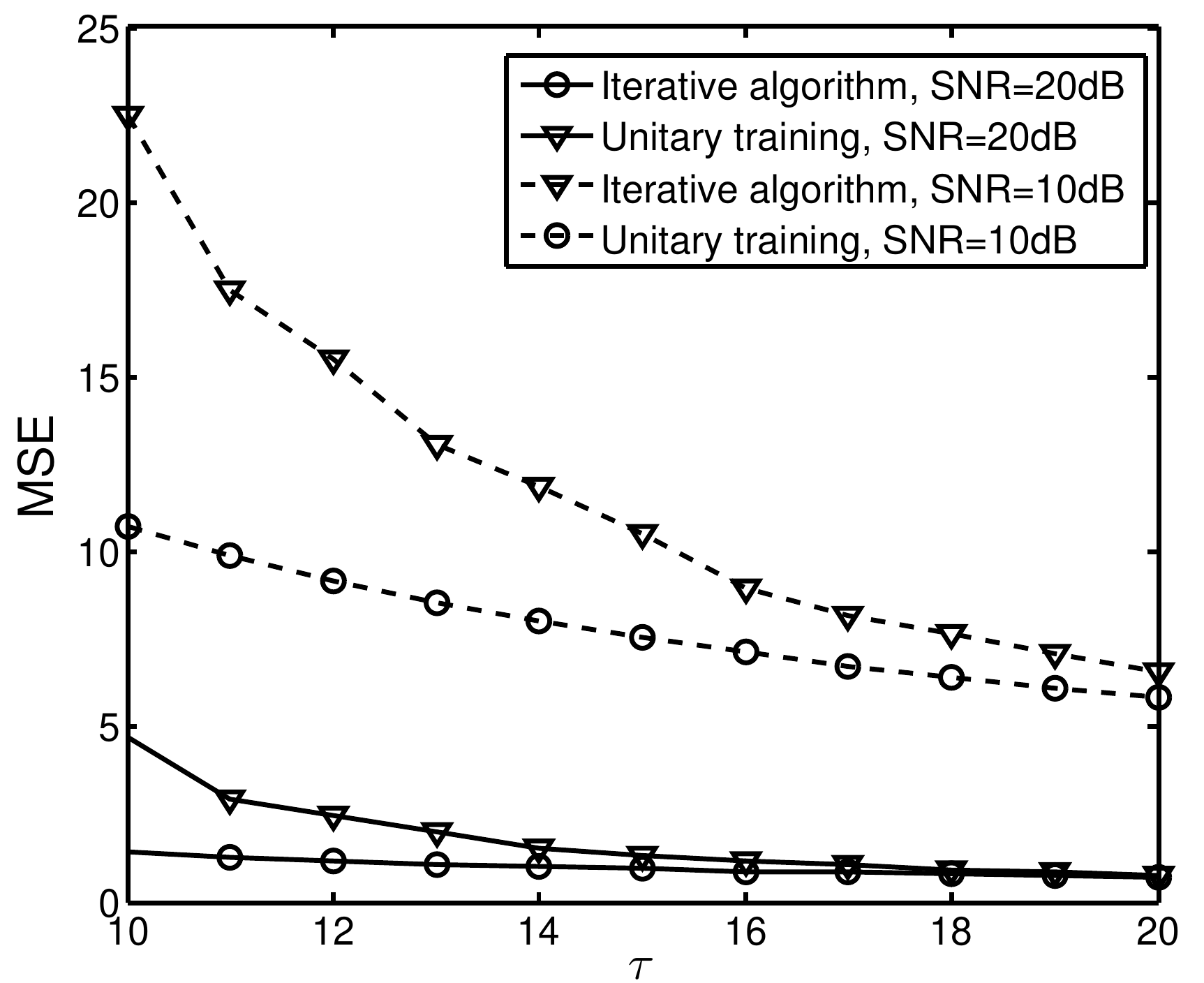}
\caption{Total mean-square-error caused by only the channel training process versus the number of training symbols of the optimal training signals given by the iterative algorithm, compared with random orthogonal training sequences. The downlink and uplink SNR are set to $20$~dB and $10$~dB, respectively. $N=8$, $M=20$. The per-user channel correlation matrices are calculated according to the one-ring model, with $D=0.5\lambda$ and $\Delta=10^\circ$.}
\label{Fig_MSE}
\end{figure}
\begin{figure}[!t]
\centering
\includegraphics[width=0.5\textwidth]{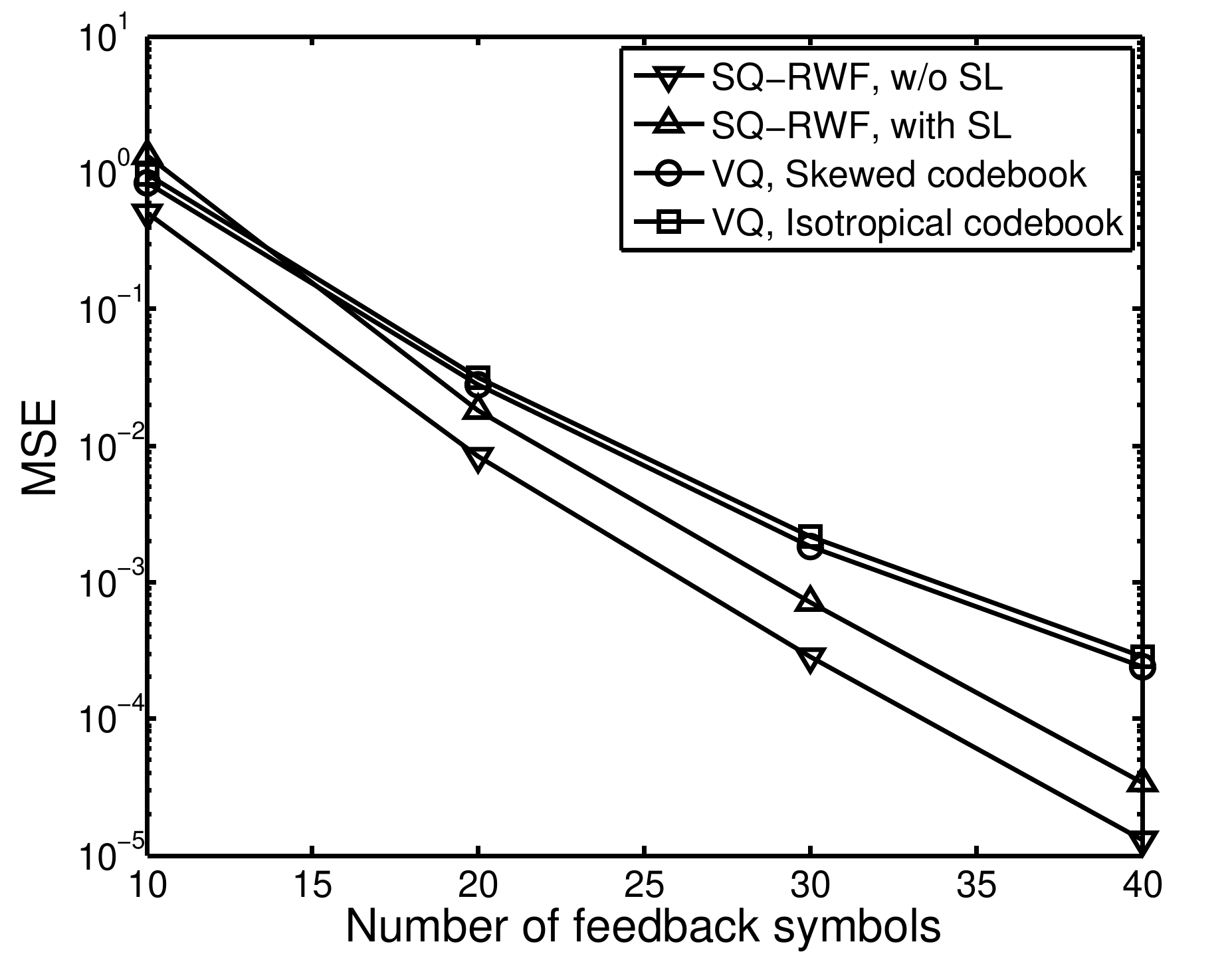}
\caption{The shaping loss (SL) is $0.75$ bits per real dimension. The downlink and uplink SNR are set to $20$~dB and $10$~dB, respectively. $N=8$, $M=20$. The per-user channel correlation matrices are calculated according to the one-ring model, with $D=0.5\lambda$ and $\Delta=10^\circ$.}
\label{Fig_SQ_VQ}
\end{figure}
\begin{figure}[!t]
\centering
\includegraphics[width=0.5\textwidth]{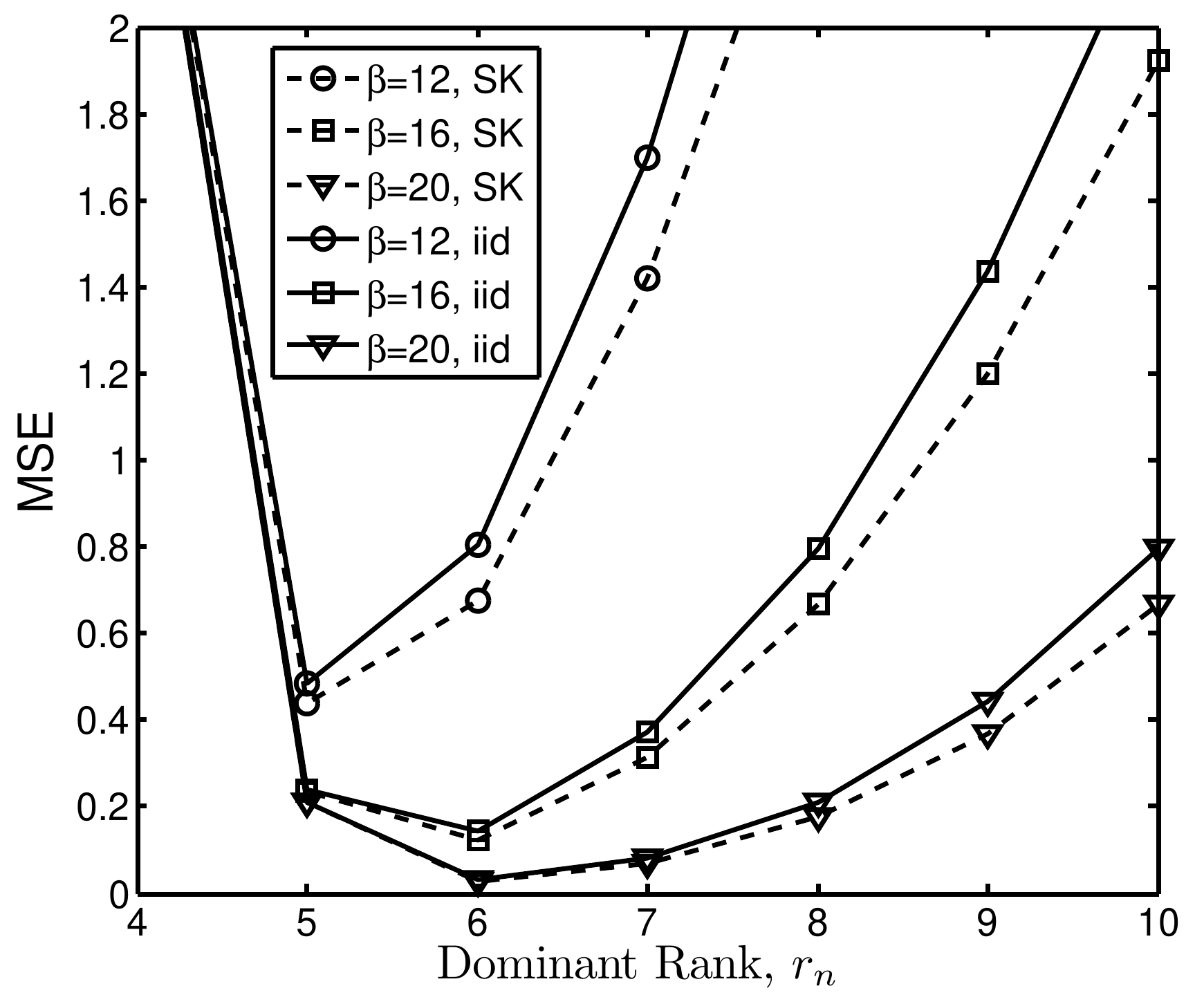}
\caption{Total mean-square-error resulting from the CSIT feedback process versus the dominant rank we choose to feedback the CSIT, with various number of feedback symbols. The downlink and uplink SNR are set to $20$~dB and $10$~dB, respectively. $N=8$, $M=20$. The per-user channel correlation matrices are calculated according to the one-ring model, with $D=0.5\lambda$ and $\Delta=10^\circ$.}
\label{Fig_EffRank}
\end{figure}
\begin{figure}[!t]
\centering
\includegraphics[width=0.5\textwidth]{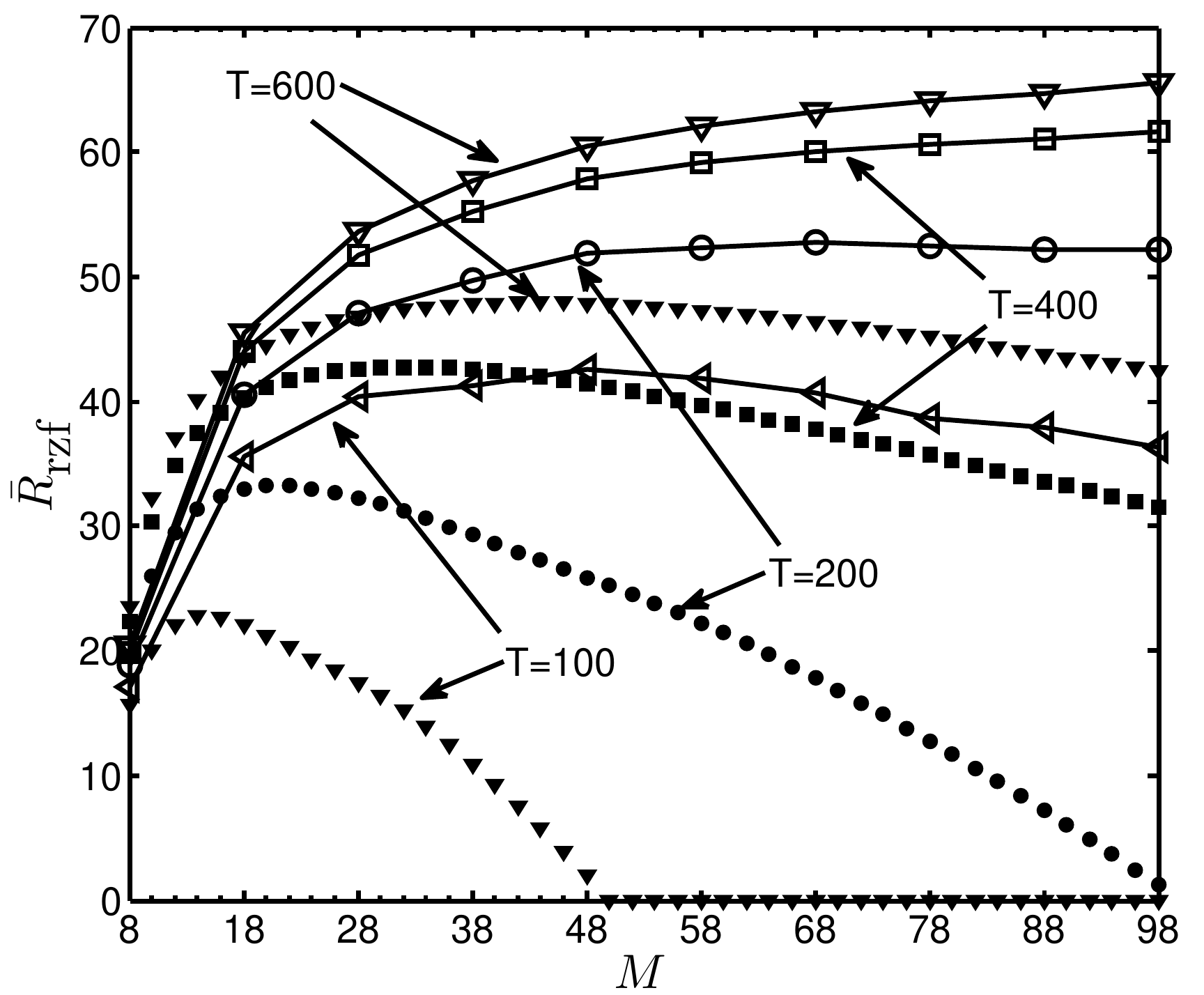}
\caption{Performance of FDD massive MIMO systems with CCMs and various block length. The downlink and uplink SNR are set to $20$~dB and $10$~dB, respectively. The number of users in the cell is $N=8$. The regularization factor of RZF precoder is $\alpha = 0.01$. The per-user channel correlation matrices are calculated according to the one-ring model, with $D=0.5\lambda$ and $\Delta=10^\circ$.}
\label{Fig_wocor}
\end{figure}
\section{Numerical Results}
\label{sec_nr}
In our simulations, we evaluate the FDD massive MIMO achievable rates with various spatially correlated channel models, and compare those with the TDD system, the FDD system with i.i.d. channels, and the JSDM scheme.
\subsection{CCMs: One-Ring Model and Laplacian Model}
First, we evaluate the eigenvalue distribution of CCMs under popular channel models. In Fig. \ref{Fig_ER}, the cumulative probability function (CDF) of the singular values of the user CCMs is shown. We adopt two models to calculate the CCM of a linear antenna array. The first one is the one-ring model \cite{Shiu00}, based on which
\begin{equation}
\label{OR}
{\left[\bm{R}\right]_{i,j}} = \frac{1}{{2\Delta }}\int_{ - \Delta  + \theta }^{\Delta  + \theta } {{e^{ - j2\pi D(i - j)\sin (\alpha )}}d\alpha },
\end{equation}
where $\Delta$ denotes the angular spread, $\theta$ denotes the azimuth angle from the user to the BS and $D$ is the spacing between two adjacent antennas. Alternatively, the Laplacian angular spectrum model is also considered \cite[Section 7.4.2]{Molisch}, where
\begin{equation}
\label{Lap}
{\left[\bm{R}\right]_{i,j}} = \frac{1}{{\sqrt 2 \Delta }}\int_{ \theta - \pi }^{\theta + \pi}  {{e^{ - \frac{{\sqrt 2 }}{\Delta }\left| {\alpha  - \theta } \right| - j2\pi D(i - j)\sin (\alpha )}}d\alpha }.
\end{equation}
From Fig. \ref{Fig_ER}, it is observed that the singular values of the CCMs are generally distributed with large deviations under various parameters, i.e., some singular values are effectively close to zeros, thus we define the number of singular values larger than a given threshold $\rho$ as the \emph{effective rank (ER)} of the CCM. Generally, as the antenna spacing is smaller, or the angular spread is smaller, the effective rank will be smaller. Note that usually the effective rank calculated by the Laplacian model is larger than that by the one-ring model, due to the one-ring model restricting the direction-of-arrival (DoA) to a finite support. Also note that the number of BS antennas is relevant, for which is shown in \cite{Adhikary13} that the ratio of ER and $M$ approaches a constant asymptotically with $M$ going to infinity. In the following simulations, we evaluate the FDD massive MIMO achievable rates under a variety of parameters and both models depicted in Fig. \ref{Fig_ER}.
\subsection{Comparison with TDD and i.i.d. FDD systems}
In presence of spatially correlated channels, the achievable rates under the proposed scheme are shown in Fig. \ref{Fig_TDDvsFDD}, in comparison with i.i.d. FDD systems and also TDD systems. The achievable rates of FDD systems with correlated channels are obtained using the training sequences obtained by the iterative algorithm in Section \ref{sec_training}, and the KLSQ feedback codebook design in Section \ref{Sec_feedback}. First, it is noteworthy that in FDD systems, in general, the achievable sum rate is not monotonously increasing with the number of BS antennas, as it does so in the TDD system, due to the fact that when the number of BS antennas grows large with FDD mode, the channel estimation dimensionality loss will become non-negligible. Therefore, there is a large rate gap between the i.i.d. FDD system and the TDD system, rendering the FDD mode unfavorable for massive MIMO transmission.

Nevertheless, when the channel is spatially correlated, the FDD system achievable sum rate under per-user CCMs is significantly larger than that in i.i.d. channels, especially when the number of BS antennas is large, thanks to the judiciously designed dominant channel estimation schemes. The rate gap between the TDD mode and the FDD mode is narrowed significantly, especially when $M$ is moderate, which suggests that it is promising to exploit the large-system gain even with FDD.

Two clarifications should be made. First the achievable rates of FDD systems are even larger than TDD systems under some parameters shown in Fig. \ref{Fig_TDDvsFDD}. The phenomenon is explained by the fact that the uplink SNR is set $10$~dB lower than the downlink SNR in the corresponding simulation results, which is typical for a cellular system due to the smaller transmit power of user-terminals, rendering the TDD system performance inferior due to the imperfect \emph{uplink} channel training. Observe that when $M$ becomes larger, the TDD system sum rate will go up unbounded, eventually surpassing the FDD system. Moreover, when the uplink SNR is set to be the same as the downlink SNR, see corresponding curves, the TDD system performs better, which is as expected. Secondly, the performance with correlated channels is slightly worse than the i.i.d. channels when the number of BS antennas is small, due to the fact that the channel capacity with i.i.d. channels is larger than the one with correlated channels, regardless of the estimation overhead, and when the number of BS antennas is small, the estimation overhead is negligible compared with the channel block length.

\subsection{Comparison with JSDM}
In Fig. \ref{Fig_JSDM}, we compare the achievable sum rates obtained by the proposed eigenspace channel estimation to the JSDM scheme \cite{Adhikary13,Adhikary132}, which was the first to exploit the spatial correlation to benefit the FDD massive MIMO system. Note that the uplink CSIT feedback is not treated in the previous JSDM papers \cite{Adhikary13,Adhikary14}. To make fair comparison, we assume that the JSDM scheme uses an isotropical VQ feedback codebook, since it is unknown whether the JSDM scheme can also benefit from a better-designed codebook for correlated channels after the pre-projection of channel vectors. To get more insights and understand the simulation results better, it is important to first illustrate the merits and demerits of the JSDM scheme compared to our scheme.

The JSDM scheme has the advantage to better suppress the channel estimation overhead. Specifically, by grouping the users based on their respective CCMs and performing the pre-beamforming, the equivalent number of BS antennas in each \emph{virtual sector}, i.e., $b_g$ in \cite{Adhikary13}, can be optimized to strike a good balance between the power gain, which scales with $b_g$, and the channel estimation overhead. In an extreme case, $b_g$ can be made as small as the number of users in each virtual sector, thus, the overall channel estimation overhead scales with the number of users in each virtual sector, which drastically decreases the dimensionality loss. However, on the downside, while the JSDM scheme adopts a \emph{divide-and-multiplex} approach, the division is imperfect, in the sense that the JSDM scheme suffers from the inherent residual \emph{inter-group interference (IGI)}, especially when the CCMs of the users in each group are different, rendering that the pre-beamforming cannot counteract the IGI completely. Notice that in our framework, the proposed dominant channel estimations incorporate all the user CCMs into the scheme design, which significantly mitigates the IGI. Moreover, it is noteworthy that the \emph{computational complexity} of the JSDM scheme is smaller compared with our proposed scheme, since our scheme deals with a higher dimensional channel matrix. \footnote{Possible operations on the channel matrix include inversion and SVD, depending on the precoding algorithm.}

Specifically, we follow the parameters used in the simulation in \cite[Section IV-C]{Adhikary132}. The \emph{fixed angular quantization} method is adopted to divide users into $G=8$ user-groups, where each group performs the per-group-processing. The quantization points are
\begin{equation}
\theta\in\{ -57.5^o, -41.5^o, -23^o, -7.5^o, 7.5^o, 23.5^0, 41.5^o, 57.5^o\},
\end{equation}
and the angular spread for the quantization matrices are identical with the user-angular-spread, which is specified in Fig. \ref{Fig_JSDM}. To keep the IGI under control, similarly with \cite{Adhikary132}, we further divide the user-groups into $2$ patterns, where only the users in the same pattern are scheduled simultaneously\footnote{For fair comparison, we set the number of users in the achievable sum rate of the proposed scheme to be half of the total users in the JSDM, since there are $2$ patterns.}. The user azimuth angles from the BS are generated uniformly from $[-60^o,60^o]$. The effective rank in each virtual sector, i.e. $r^\star$ in \cite{Adhikary13}, is chosen neglecting extremely small eigenvalues of the channel correlation matrix, and $b_g$ is chosen to optimize the sum rate by exhaustive search. The training sequences of each virtual sector are i.i.d. sequences as in \cite[Section VI]{Adhikary13}.

The deterministic equivalents for the JSDM scheme are computed based on \cite[Appendix A]{Adhikary13}, with generalizations to distinct CCMs within each user-group. The details are again omitted for brevity.

It is observed from Fig. \ref{Fig_JSDM} that the JSDM scheme achieves better sum rate when the channel coherence time is small, e.g., $T=100$, and the number of BS antennas $M$ is large. Qualitatively, this is expected since the small channel coherence time and large $M$ both put more weight in the need to suppress the channel estimation overhead, and based on \cite{Adhikary13}, a large $M$ also leads to the fact that the eigenvectors of the channel correlation matrices can be well approximated by the columns of a Discrete Fourier Transform (DFT) matrix, which ensures orthogonality as long as the angular of arrival (AoA) intervals of different users are disjoint. On the other hand, the achievable sum rate of our proposed eigenspace channel estimation shows evidently better rate when the channel coherence time is larger, which elevates the urgency to suppress the channel estimation overhead, or when the angular spread of users is larger, which causes larger residual IGI in the JSDM scheme. Notice that large angular spread also decreases the achievable rates of our scheme, due to the increased channel estimation dimensionality, however our scheme turns out to be more resilient in this regard. Although there are several parameters in the JSDM scheme, that may be properly tuned to achieve better rate than the eigenspace channel estimation scheme, the eigenspace channel estimation scheme still has the advantage of low complexity, and the optimization for the JSDM scheme goes out of the scope of this paper.

\subsection{Performance Gain Leveraging Eigenspace Channel Estimation}
\label{sec_TF}
Furthermore, we demonstrate how much gain we can get from leveraging the training and feedback schemes designed for the multi-user CCMs, by comparing with using the unitary training and feedback schemes as in the i.i.d. channel case. The rate gain is depicted in Fig. \ref{Fig_OPTvsORG}, showing leveraging eigenspace channel estimation can indeed improve the sum rate of FDD massive MIMO system with spatially correlated channels. The detailed performance analyses of eigenspace channel training and feedback are shown in Fig. \ref{Fig_MSE} and Fig. \ref{Fig_EffRank}, respectively.

For the training process, the MSE performance of the iterative algorithm we developed in Section \ref{sec_training}, which finds the optimal training sequence with per-user CCMs, is shown in Fig. \ref{Fig_MSE}. When the number of training symbols is small, the total MSE achieved by the iterative algorithm is much lower than the orthogonal training sequences, due to the fact that in presence of channel correlation, the training sequences obtained by our algorithm can find the eigenspace that needs to be estimated more accurately and concentrate the power to that subspace. Note that when the downlink SNR is large and the number of training symbols is large enough\footnote{In this case $\tau \ge M$, such that we have enough number of channel observations to recover the channel coefficients perfectly when SNR goes to infinity} to train all the subspaces, the unitary training sequences
are asymptotically optimal. Such observations are further evaluated by setting the downlink SNR to $10$~dB, which shows a certain MSE gap between the optimal training sequences and the unitary training sequences, even when $\tau=M$.

Fig. \ref{Fig_SQ_VQ} shows the total MSE performances of KLSQ and VQ with isotropical and skewed codebooks. It is observed that KLSQ achieves better MSE performance, even with the shaping loss when the number of feedback symbols is large. Notice that in general VQ is more efficient than SQ, especially when the vector is correlated. However, after the KL-transform, the channel vector is decorrelated into independent Gaussian variables with non-identical variances, in which case the RWF is the optimal bit allocation in terms of MSE distortion.

The impact of the dominant rank, i.e., $r_n$, we choose in the VQ feedback process on the MSE is shown in Fig. \ref{Fig_EffRank}, with different number of feedback symbols. The tradeoff between the quantization accuracy of the effective channel and the estimation error resulting from the neglected eigenspace of the CCM is shown. It is observed that there exists an optimal number of $r_n$ in terms of minimizing the total feedback error. The optimal $r_n$ is increasing with the number of feedback symbols, for the reason that when we have more feedback symbols, we can afford to estimate a higher-dimensional eigenspace, rendering a better accuracy of the CSIT feedback estimation. The performance of the skewed feedback codebook is also shown in the figure. The gain in terms of MSE is fairly small, when the optimal dominant rank is chosen, because the error mainly stems from neglecting the non-dominant eigenspace. Note that when $r_n$ is large, the performance gain of the skewed codebook is more evident since the MSE in this regime is dominated by the channel quantization. It is worth mentioning that the absolute values of the feedback MSE are fairly small, compared with the training error. We find that the channel estimation error mainly comes from the analog downlink channel training process, rather than the digital feedback process, for the reason that the MSE scales inversely with the number of training symbols \eqref{MSE}, but exponentially with the feedback symbols \eqref{h_quant_error_cov}.

In Fig. \ref{Fig_wocor}, the achievable sum rate improvement with spatially correlated channels is shown, under various values of block length. The block length characterizes how long the channels stay constant, both temporally and spectrally. Significant rate improvement, which is up to two-fold, is found from the figure. The results suggest that under the spatially correlated channels, which is especially common with mm-wave channels \cite{Adhikary14}, along with a well-designed transmission strategy, namely the training and feedback schemes, the FDD system is capable of realizing significant massive MIMO gain.

\section{Conclusions}
\label{sec_cl}
By computing the achievable rates with a RZF precoder of FDD massive MIMO systems, on account of the downlink channel training and uplink CSIT feedback dimensionality loss and corresponding channel estimation error, we showed that spatial channel correlation at the BS side is beneficial to the FDD massive MIMO system. The benefit is especially prominent if the channels are strongly correlated, namely the CCMs are effectively rank-deficient. In particular, we propose an iterative algorithm to find the optimal channel training sequences in presence of multiuser spatial channel correlation, and a KL-transform followed by SQ with RWF bit-loading feedback codebook design, which is extremely computationally efficient and thus easy to implement in practice while achieving near-optimal performance. Our proposed approach, which achieves dimensionality reduction channel estimation even without channel pre-projection, can be seen as an alternative to the projection and effective channel approach in the JSDM scheme. Moreover, it is noteworthy that while achieving a significant performance gain, our approach only requires minimal modifications of the widely-used training-based transmission scheme, and thus it is easy to implement.

Numerical results show significant rate improvements when leveraging our proposed eigenspace channel estimation approaches under spatially correlated channels, in comparison with i.i.d. FDD massive MIMO systems. In fact, when the channel correlation is strong and the number of BS antennas is not very large, the achievable sum rate of FDD massive MIMO systems can even outperforms TDD systems. Comparisons with the JSDM scheme reveal both schemes have advantages under different channel conditions, such as coherence time and angular spread. In particular, our proposed schemes display better performance when channel coherence time is large, or the angular spread of the users is large, while requiring a higher computational complexity due to operating on a higher-dimensional matrix.

These results suggest that in FDD massive MIMO systems, increasing spatial channel correlation, e.g., by decreasing the antenna spacing, more line-of-sight transmission, etc., can be beneficial. Although this differs from the favorable propagation conditions in TDD systems, which prefer i.i.d. channels to maximize the total DoF, the FDD system benefits significantly from correlation, which enables dimensionality reduction as far as channel estimation is concerned. The tradeoff between the DoF of the downlink BC and the spatial correlation in FDD massive MIMO is an interesting problem for future work.
\appendices
\section{Proof of Theorem \ref{thm_mseq}}
\label{App_thm_mseq}
\begin{IEEEproof}
The MSE of the skewed codebook is expressed as
\begin{equation}
\label{app1}
\textrm{MSE}_\textrm{q} = \mathbb{E}_{\bm{z}_n} \left\{ \mathbb{E}_{\mathfrak{C}_{\textrm{sk}}} \left[ \bm{z}_n^\dag \bm{\Lambda}_n \bm{z}_n - \max_i \left[\frac{\bm{f}_i^\dag \bm{\Lambda}_n \bm{z}_n \bm{z}_n^\dag \bm{\Lambda}_n \bm{f}_i }{\bm{f}_i^\dag \bm{\Lambda}_n \bm{f}_i}\right] \right]\right\}+ \textrm{tr} \left[ \bm{\bar U}_n \bm{\bar \Sigma}_n \bm{\bar U}_n^\dag \right],
\end{equation}
where $\bm{h}_n = \bm{Q}_n \bm{\Lambda}_n^{\frac{1}{2}}\bm{z}_n$, $\bm{Q}_n$ is a unitary matrix, $\bm{z}_n \sim \mathcal{CN}(\bm{0},\bm{I}_{r_n})$. Define the first term in \eqref{app1} as $\Delta_1$. We obtain
\begin{IEEEeqnarray}{rcl}
\label{app2}
\Delta_1 &=& \mathbb{E}_{\bm{z}_n} \left\{\int_{0}^{\bm{z}_n^\dag \bm{\Lambda}_n \bm{z}_n} {\left[\textrm{Pr}\left( \frac{\bm{f}_i^\dag \bm{\Lambda}_n \bm{z}_n \bm{z}_n^\dag \bm{\Lambda}_n \bm{f}_i }{\bm{f}_i^\dag \bm{\Lambda}_n \bm{f}_i} \le x | \bm{f}_i^\dag \bm{f}_i = 1\right)\right]^N dx}\right\} \\
\label{app3}
&\le& \mathbb{E}_{\bm{z}_n} \left\{\int_{0}^{\bm{z}_n^\dag \bm{\Lambda}_n \bm{z}_n} {\left[\textrm{Pr}\left( \frac{\bm{f}_i^\dag \bm{\Lambda}_n \bm{z}_n \bm{z}_n^\dag \bm{\Lambda}_n \bm{f}_i }{\lambda_1} \le x | \bm{f}_i^\dag \bm{f}_i = 1\right)\right]^N dx}\right\} \\
\label{app4}
&=& \mathbb{E}_{\bm{z}_n} \left\{ \int_{0}^{\bm{z}_n^\dag \bm{\Lambda}_n \bm{z}_n} {\left[\textrm{Pr}\left( \left(\bm{f}_i^\dag \frac{\bm{\Lambda}_n \bm{z}_n}{\sqrt{\bm{z}_n^\dag\bm{\Lambda}_n^2 \bm{z}_n}}\right)^2 \le \frac{\lambda_1 x}{\bm{z}_n^\dag\bm{\Lambda}_n^2 \bm{z}_n} | \bm{f}_i^\dag \bm{f}_i = 1\right)\right]^N dx} \right\} \\
\label{app5}
&\approx& \mathbb{E}_{\bm{z}_n} \left\{ \int_{0}^{\frac{\bm{z}_n^\dag \bm{\Lambda}_n^2 \bm{z}_n}{\lambda_1}} {\left[\textrm{Pr}\left( \left(\bm{f}_i^\dag \frac{\bm{\Lambda}_n \bm{z}_n}{\sqrt{\bm{z}_n^\dag\bm{\Lambda}_n^2 \bm{z}_n}}\right)^2 \le \frac{\lambda_1 x}{\bm{z}_n^\dag\bm{\Lambda}_n^2 \bm{z}_n} | \bm{f}_i^\dag \bm{f}_i = 1\right)\right]^N dx} \right\} \\
\label{app6}
&=& \mathbb{E}_{\bm{z}_n} \left\{ \frac{\bm{z}_n^\dag \bm{\Lambda}_n^2 \bm{z}_n}{\lambda_1^{(n)}} \int_{0}^{1} {\left[1-\left(1-x\right)^{r_n-1}\right]^N dx} \right\} \\
\label{app7}
&\approx& \frac{{\sum\limits_{i = 1}^{{r_n}} {(\lambda _i^{(n)})^2} }}{{{\lambda _1^{(n)}}}}{2^{\frac{{ - {B_n}}}{{{r_n} - 1}}}},
\end{IEEEeqnarray}
wherein the equality \eqref{app2} follows from integrating \eqref{app1} by parts, the approximation in \eqref{app5} follows from the work in \cite[Appendix J]{Raghavan13}, which shows the dominant term of the integral in \eqref{app4} is \eqref{app5}, then by \cite[Corollary 1]{Chun07}, \eqref{app6} and \eqref{app7} follows.
\end{IEEEproof}
\bibliographystyle{ieeetr}
\bibliography{AR}
\end{document}